\documentclass[a4paper,12pt]{article}

\usepackage{geometry}       
\usepackage{graphicx}       
\usepackage{subfig}         
\usepackage{amsmath}
\usepackage{float}
\usepackage{mathptmx}       
\newgeometry{vmargin={25.4mm}, hmargin={27mm,27mm}}
\usepackage[colorlinks]{hyperref}
\usepackage{mwe}
\hypersetup{colorlinks,linkcolor={black},citecolor={black},urlcolor={black}} 

\usepackage[nameinlink]{cleveref}
\Crefname{figure}{Fig.}{Figs.}

\usepackage[
backend=biber,
style=numeric,
sorting=none
]{biblatex}
\usepackage{authblk} 
\title{A Toolbox for Optimization of Reconstruction and Post-processing Pipelines in In Vivo $^{31}$P MR Imaging }

\author[1,2,4]{Pontus Pandurevic}
\author[1,2]{Mark Stephan Widmaier}
\author[1,2]{Zhiwei Huang}
\author[1,2,3]{Lijing Xin}

\affil[1]{Center for Biomedical Imaging (CIBM), Switzerland}
\affil[2]{Animal Imaging and Technology, Ecole Polytechnique Fédérale de Lausanne (EPFL), Lausanne, Switzerland}
\affil[3]{Institute of Physics (IPHYS), Ecole Polytechnique Fédérale de Lausanne (EPFL), Lausanne, Switzerland}
\affil[4]{Department of Physics, KTH Royal Institute of Technology, Stockholm, Sweden}

\date{}

\begin{document}

\maketitle

\newpage
\section{Abstract}

BACKGROUND: Low signal-to-noise ratio (SNR) is a persistent problem in X-nuclei imaging due to lower gyromagnetic ratios, and tissue levels compared to hydrogen. To address the SNR challenge, strategies including the use of high-fields, advanced coil, sequence design, as well as reconstruction and processing methods are commonly applied. An optimized dedicated reconstruction and processing pipeline is essential to achieve the highest SNR for X-nuclei imaging. 

PURPOSE: This work aims to focus on developing a toolbox that includes various reconstruction and processing pipelines for fast $^{31}$P MR imaging, to evaluate and determine the optimal pipeline for two types of fast phosphorus imaging data, i.e., $^{31}$P -GRE and bSSFP-like MRF imaging.

Methods: The toolbox includes coil combination methods (adaptive coil combination and whitened-SVD), k-space filtering, reconstruction methods (Kaiser-Bessel regridding (KB) with FFT and non-uniform FFT), and denoising methods (compressed sensing and MP-PCA).  $^{31}$P MR imaging datasets were acquired at 7T using fast imaging sequence based on a stack of spiral encoding with a double-tuned (Tx/Rx) birdcage coil and a 32-channel receive ($^{31}$P) phased array coil. The GRE acquisition parameters were: readout duration of 18.62ms, flip angle (FA) of 59°, repetition time (TR) of 5400ms, and echo time (TE) of 0.3ms. The MRF acquisition used a bSSFP-like scheme with a readout duration of 9.12ms, TR of 19.8ms, and TE of 0.3ms.

RESULTS: SNR maps show that, for GRE data, adaptive coil combination with Kaiser-Bessel regridding (AC-KB) outperformed all other methods, achieving SNRs of 38 and 3.4 for phantom and in vivo reconstructions, respectively, compared to the second-best performers with SNRs of 17 and 2.4. For MRF data, wSVD with KB achieved an SNR of 18, outperforming the second best (AC-KB) with SNR of 16.

CONCLUSION: In conclusion, we demonstrated a toolbox for optimizing reconstruction pipelines in fast $^{31}$P MR imaging, revealing that the optimal methods are highly dependent on acquisition type and data characteristics. Kaiser-Bessel regridding outperforms NUFFT for reconstruction; adaptive coil combination is preferred for GRE data, while whitened-SVD is superior for MRF data. MP-PCA denoising effectively enhances GRE images, whereas compressed sensing is more suitable for MRF data. These findings underscore the necessity of carefully selecting processing methods for X-nuclei imaging, as conventional hydrogen MRI techniques may not be optimal. Future investigations should aim to determine the best pipelines for other X-nuclei datasets using the implemented toolbox.

Keywords: X-Nuclei, Phosphorous imaging, GRE, MRF, Reconstruction, Denoising, Coil Combination

\newpage

\tableofcontents

\newpage

\section{Introduction}
$^{1}$H MRI is a common imaging modality that provides structural and functional information about living organs. In contrast, non-hydrogen nuclei (X-nuclei), such as \( ^{23}\text{Na} \), \( ^{31}\text{P} \) and $^{17}$O, offer diverse insights into  physiological processes in vivo. Among these, $^{31}$P metabolites are mainly involved in bioenergetics, and are implicated in neurodegenerative diseases \cite{xnucl}, as well as other pathologies such as cancer \cite{cancer} and psychiatric disorders \cite{bipolarity}.
\newline
\newline
The main challenge in X-nuclei imaging and spectroscopy is the low signal-to-noise ratio (SNR) which leads to prolonged scanning times. The lower gyro-magnetic ratio of the nuclei compared to $^{1}$H and their lower tissue levels relative to $^{1}$H in the water, result in a lower relative SNR (to $^{1}$H MRI) of approximately  $10^{-6}$ for $^{31}$P. To address the SNR challenge, strategies including the use of high magnetic fields, advanced coil such as phased array receivers, efficient sequence design, and fast k-space sampling strategies, as well as reconstruction and processing methods such as denoising approaches can be applied. 
\newline
\newline
Phased array coils offer higher SNR relative to a birdcage receiver coil.  The actual gain in SNR achieved depends not only on the coil itself, but also on the coil combination method. The solution of combining signals from phased arrays for 1H MRI has been well addressed, while it is challenging for X-nuclei data as they usually has low SNR signal from individual coil elements. Simple way to combine the signal from each coil is the  "sum of squares" (SoS). However, it only provides the magnitude information of the data. More advanced combination methods have been proposed to access the phase information and increase the SNR. These methods utilize assumptions about the noise, either in a spectral manner, such as whitened singular value decomposition (wSVD), which assumes that the noise from the different spectra recorded by different coils is correlated \cite{wsvd}, or in the image domain, such as adaptive coil combination (AC), which estimates an optimal filter based on averaging matrix cross-products over individual coil images and then applying this filter pixel-wise on the coil images \cite{AC}. Both AC and wSVD have been shown to increase SNR compared to SoS \cite{AC1lit, usesWSVD1, mostlikely}, with a comparison between them for the detection of breast cancer tumors showing AC outperforming wSVD \cite{mostlikely++}.
\newline
\newline
Fast spatial encoding methods, especially non-uniform trajectories, are commonly used for X-nuclei with a denser sampling in the center to increase SNR. Non-uniform trajectories however need to be mapped to a Cartesian k-space before transforming it or directly transform it to the image space. Two commonly used algorithm-based methods are Kaiser-Bessel regridding (KB), as proposed in \cite{gridKB}, which convolves the k-space data with a gridding kernel and then samples it onto a uniform space, and the non-uniform fast Fourier transform (NUFFT) with KB kernels \cite{nufftOG}. NUFFT has mostly replaced regridding algorithms as it outperforms KB as shown for different MRI applications\cite{KBbetterNot, nufftBetter}. However, NUFFT demands more computational time\cite{KBbetterNot, nufftBetter}. Both regridding and NUFFT have been used in multiple sodium imaging studies\cite{sodiumNufft1, sodiumRecon2} and NUFFT has grown into the standard method for MRF studies\cite{Ma2013,mrf2, KratzerMRF}. New deep learning (DL) methods have been investigated and showed promise\cite{DLMRF}, however they rely on sufficient training data often lacking in X-Nuclei applications. While the classical methods have been intensively investigated for $^{1}H$ MRI, a similar comparison for X-Nuclei does not, to the best of the authors knowledge, exist. Therefore, the comparison of classical methods such as NUFFT and KB in X-nuclei applications, is prioritized over more novel methods such as DL. 
\newline
\newline
To further improve SNR in the processing step, denoising methods are commonly used. We implemented and examined two denoising techniques:the principal component based method, namely the Marcenko-Pasteur principal component analysis (MP-PCA) and the widely used iterative method, compressed sensing (CS). Note that while it has been shown that deep learning approaches \cite{DL1, DL2} perform well for denoising, a lack of data in X-nuclei applications hinders its application. CS, as proposed in \cite{Lustig2007}, is shown to be effective and robust in hydrogen imaging\cite{CSFDA}, and works by incorporating the first and second order derivatives along with the L2 norm in the residual, thus preserving edges in the images while being translationaly and rotationaly invariant \cite{cs}. Beyond hydrogen imaging, its efficiency has also been shown in x-nuclei imaging\cite{CSsodium1,cssodium2}, as well as spectroscopic imaging (MRSI). MP-PCA, along with other locally low-rank constraints, has shown to be effective in denoising and accelerating acquisition \cite{LLR, MP-PCA}. It separates a high-dimensional data into significant signal and noise eigenvalues, where the cut-off for signals is determined by the Marcenko-Pasteur distribution. This identifies the significant principal components, which are likely to contain the true signal, separate from those that are mostly noise from thermal fluctuations \cite{MP1, MP2}.
\newline
\newline

Since X-nuclei data differs significantly from proton MRI data, mainly due to its lower SNR and spatial resolution, the commonly used reconstruction and processing methods for 1H MRI remain to be reevaluated according to the dedicated X-nuclei datasets. This work aims to implement a toolbox that includes various reconstruction and processing pipelines for fast $^{31}$P MR imaging, to evaluate and determine the optimal pipeline for two types of fast phosphorus imaging data, i.e., $^{31}$P -GRE and bSSFP-like MRF imaging. The implemented pipeline in the toolbox contains the combination of coil combination methods (adaptive coil combination and whitened-SVD), k-space filtering, reconstruction methods (Kaiser-Bessel regridding (KB) with FFT and non-uniform FFT), and denoising methods (compressed sensing and MP-PCA). This pipeline will also facilitate further studies to determine an optimal pipeline for other X-nuclei datasets.
\newpage

\section{Methods}

\subsection{$^{31}$P MR data acquisition}
MR experiments were conducted on a Siemens Terra X 7T/80 cm MR scanner (Siemens Healthineers, Erlangen, Germany) using a double-tuned (Tx/Rx) birdcage coil along with a 32-channel receive ($^{31}$P) phased array coil (RAPID Biomedical, Rimpar, Germany). Data were acquired from a spherical phantom with a diameter of 17 cm, filled with 50 mM Pi solution (Carl Roth GmbH \& Co. KG, Karlsruhe, Germany). and two healthy male participants (18 and 45 years old) who provided written informed consent. Following a localizer image, a shim volume covering the whole brain was placed, and 3D-volume shimming (Siemens GRE brain) with intermittent frequency adjustment was applied. The 3D $^{31}$P GRE acquisition parameters were: a stack of spiral readout with a duration of 18.62 ms, 8ms frequency selective excitation pulse (carrier frequency on phosphocreatine (PCr)) with a flip angle (FA) of 59°, repetition time (TR) 5400 ms, echo time (TE) 0.3 ms\cite{markB1}, matrix size of 32 × 32 × 11, and field of view (FOV) of 230 × 230 × 220 mm$^{3}$. The 3D $^{31}$P MRF acquisition\cite{markismrm} used a bSSFP-like sequence with a spiral readout (9.12 ms), TR of 19.8 ms, TE of 0.3 ms, matrix size of 32 × 32 × 11, and FOV of 230 × 230 × 220 mm$^{3}$. Both sequences were based on a stack of spirals.

\subsection{Reconstruction}
All data were processed using a custom toolbox in MATLAB (The MathWorks, Inc., Natick, Massachusetts, USA). For the GRE data, the resulting dimension was [k-space points × slices], and for the MRF data, [k-space points × slices × flip angle (FA) index]. The data were Fourier transformed along the slice dimension prior to reconstruction. Image reconstruction was performed using either a Kaiser-Bessel (KB) gridding algorithm to convert the k-space data to a Cartesian grid before applying a 2D Fast Fourier Transform (FFT), or using Non-Uniform Fast Fourier Transform (NUFFT). The density compensation function required for both reconstructions was based on the Voronoi diagram \cite{36,37,38}. For the in vivo GRE experiments, the spiral trajectory was truncated into 100\% (1782 spiral sampling points), 66\% (1434 sampling points), and 33\% (861 sampling points) of k-space, effectively filtering it by reducing its length. The MRF sequence inherently uses a 33\% k-space trajectory in the sequence design. When NUFFT was used for reconstruction, the images were zero-filled in the k-space dimension to the full spiral length (100\%) of 1782 sampling points.

\subsection{Coil combination}

Measurements were recorded using a 32-channel phased receiver arrays, resulting in 32 distinct yet correlated signals. The application of coil combination varied depending on the method and data type used.
\\
\\
For GRE data, the weighted singular value decomposition (wSVD) was computed for each slice in the time domain, similar to the single-voxel spectroscopy approach, treating the spiral readout as a free induction decay (FID). For MRF data, the data were first reconstructed, and then the wSVD coil combination was computed along the FA dimension, processing each pixel in the reconstructed images.
\\
\\
When using adaptive combine (AC), it was applied in the image domain. For MRF data, the required coil sensitivity maps (CSMs) were computed from an averaged image along the FA index for both phosphocreatine (PCr) and adenosine triphosphate (ATP) signals, omitting the first 100 FAs of the PCr signal, which contain the inversion recovery profile of PCr in the MRF sequence. These CSMs were then used during the AC process for images of individual FA.
\\
\\
Using the sum of squares (SoS) method, the images were first reconstructed, and then the coils were combined by summing in the image domain, reducing the coil dimension from 32 to 1.

\subsection{Denoising}
\begin{figure}[H]
    \centering
    \includegraphics[scale=0.42]{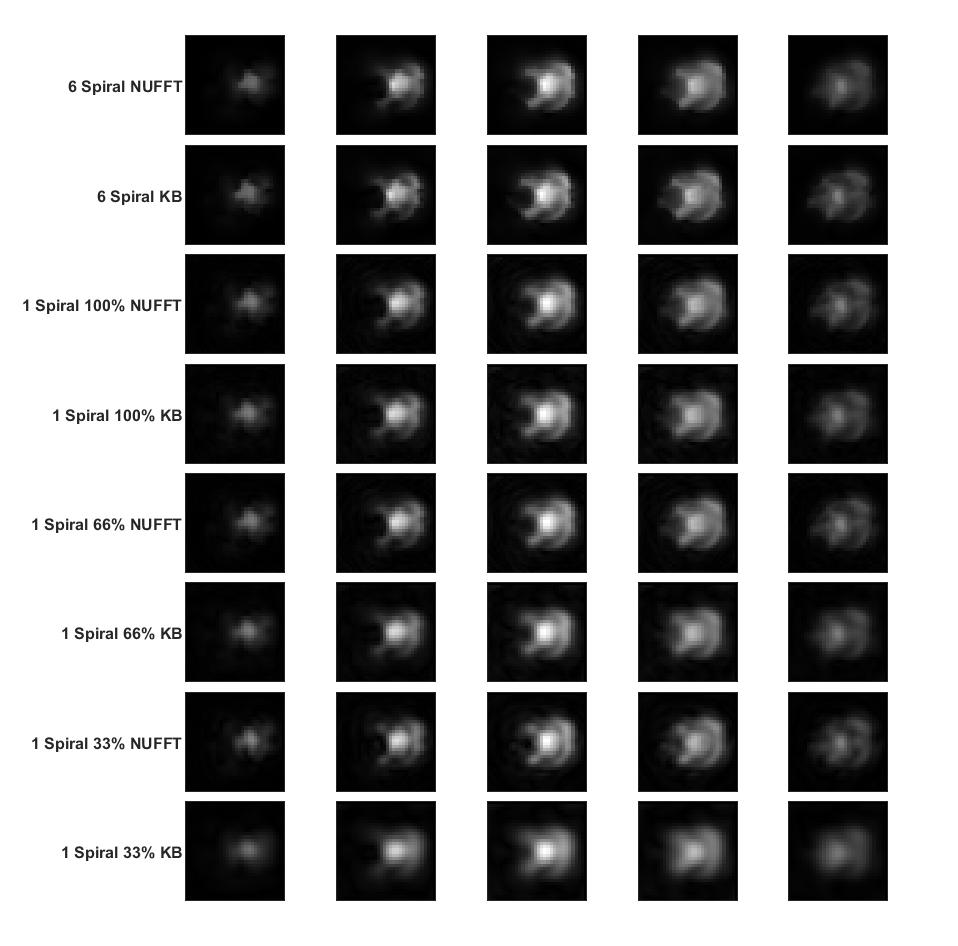}
    \caption{1H GRE MRI images reconstructed with NUFFT or KB using 6 spirals per slice or 1 while also decreasing the k-space sampling from 100\%, 66\%, and 33\%, with CS applied with a 0.001 regularization parameter. This data was acquired with the 1H birdcage TX/RX coil (single channel) outside of the 32 ch 31P phased arrays.}
    \label{1HNoNoiseCS0.001}
\end{figure}

\begin{figure}[H]
    \centering
    \includegraphics[scale=0.42]{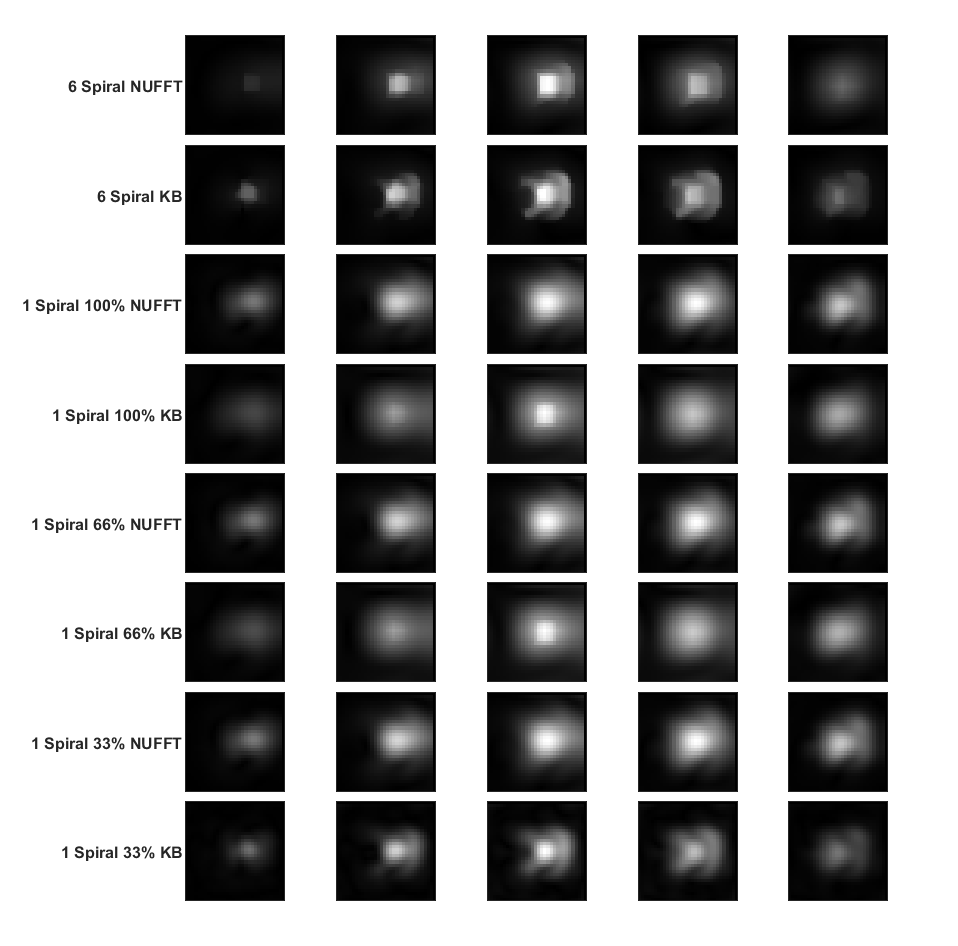}
    \caption{1H GRE MRI images reconstructed with NUFFT or KB using 6 spirals per slice or 1 while also decreasing the k-space sampling from 100\%, 66\%, and 33\%, with CS applied with a 0.01 regularization parameter}
    \label{1HNoNoiseCS0.01}
\end{figure}

\begin{figure}[H]
    \centering
    \includegraphics[scale=0.42]{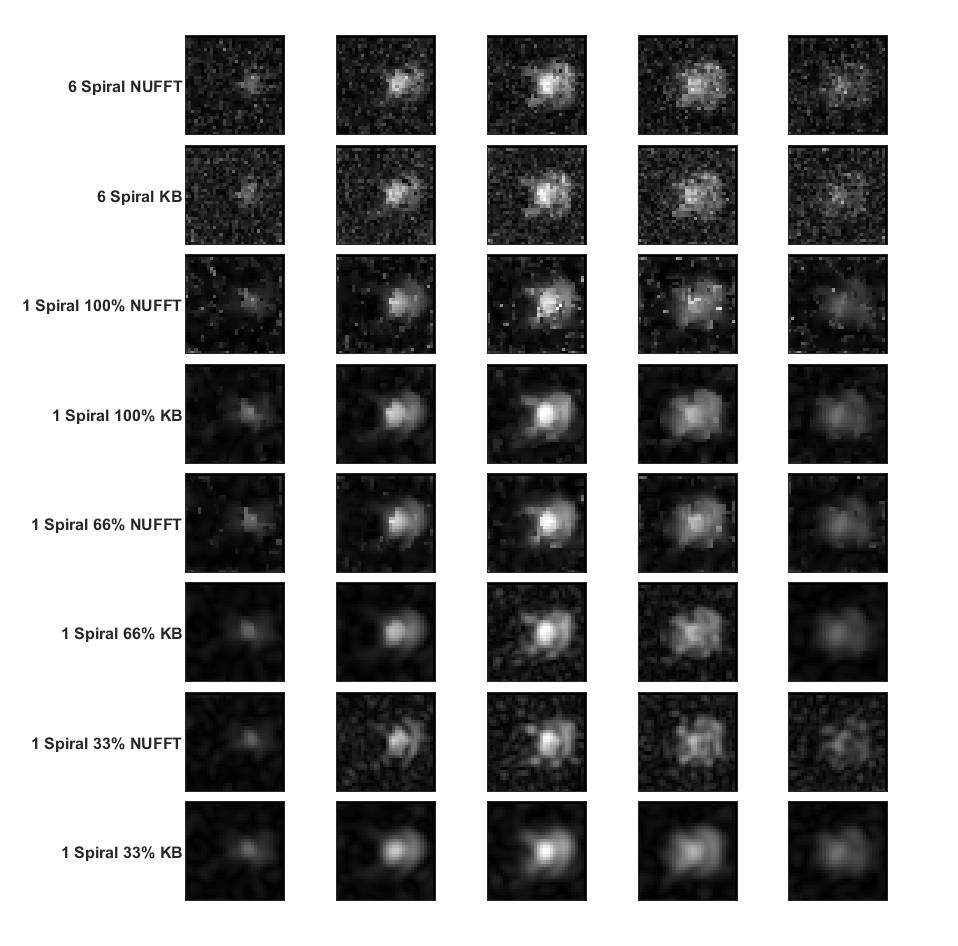}
    \caption{1H GRE MRI images reconstructed with NUFFT or KB using 6 spirals per slice or 1 while also decreasing the k-space sampling from 100\%, 66\%, and 33\%, with added Gaussian noise and CS applied with a 0.001 regularization parameter}
    \label{1HNoiseCS0.001}
\end{figure}

\begin{figure}[H]
    \centering
    \includegraphics[scale=0.42]{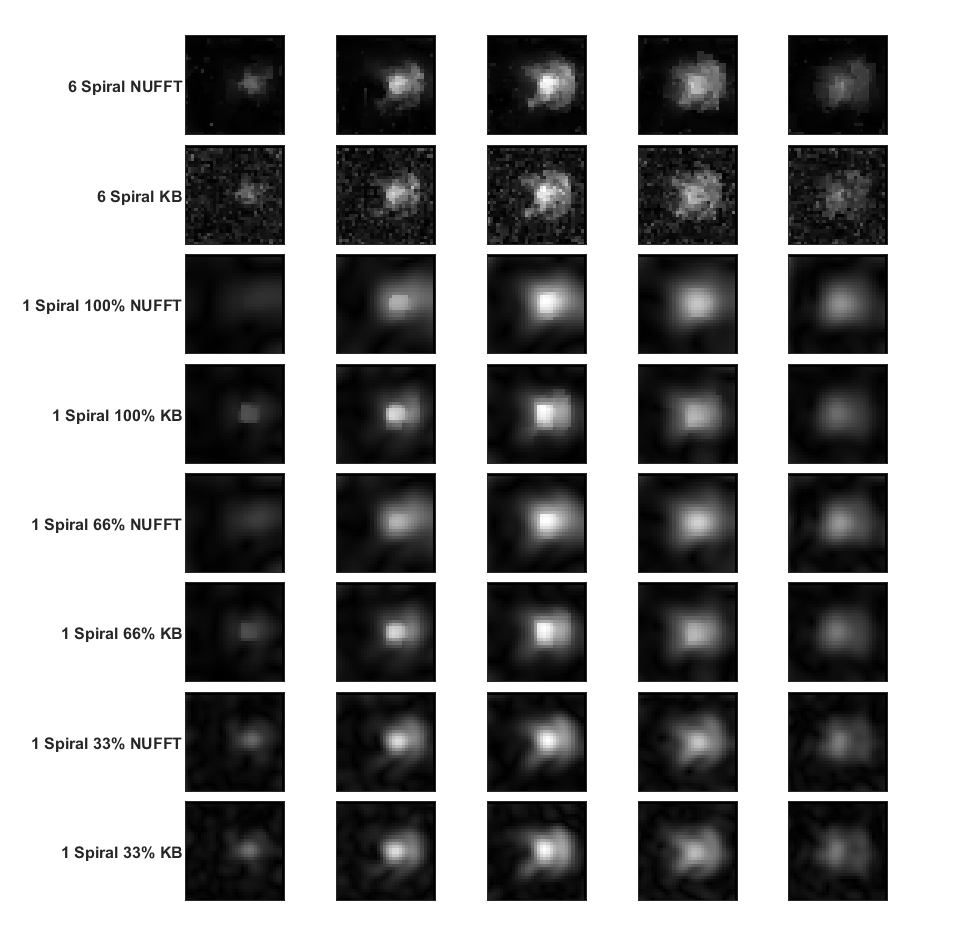}
    \caption{Hydrogen data reconstructed with NUFFT or KB using 6 spirals per slice or 1 while also decreasing the k-space sampling from 100\%, 66\%, and 33\%, with added Gaussian noise and CS applied with a 0.003 regularization parameter}
    \label{1HNoiseCS0.003}
\end{figure}

Two denoising methods were implemented in this toolbox: CS and P-PCA. CS was always applied as a final step on the 32 × 32 × 11 images slice by slice.
\\
\\
For the MRF data, MP-PCA\cite{MP1, MP2} was applied in two stages. First, on the raw data, where a 1D FFT was performed along the slice dimension, and then the data were split into real and imaginary parts before being concatenated in the time-point dimension. Then MP-PCA was applied on a data matrix with time-points × (slices × FAs) . Second, MP-PCA was applied to the images after reconstruction and coil combination. The real and imaginary parts of the images were concatenated before applying MP-PCA. For the GRE experiments, MP-PCA was applied only on the images after reconstruction.
\\
\\
For CS the completed images were flattened to dimension Nres$^{2}$, before TGV denoising was applied slice by slice as described in algorithm 1 in \cite{cs}. For the regularization parameter $\lambda$ a search was made to find the optimal value (Supplementary files Figures \ref{1HNoNoiseCS0.001}-\ref{1HNoiseCS0.003})

\subsection{Signal-to-noise ratio (SNR) maps.}
To evaluate the performance of tested pipeline combinations, the results are presented as SNR maps, where the SNR was calculated according to Equation (\ref{eq:snr}). The values for $\sigma$ and $\mu_{\text{Noise}}$ were calculated using MATLAB's \texttt{mean} and \texttt{std} functions from a pure noise region. For SNR quantification, the mean and standard deviation from a high-signal region over three slices were calculated. The specific regions used are shown in the appendix in Figures \ref{noisePhantom}-\ref{NoisePCR}.

\begin{equation}
    \text{SNR} = \frac{\text{Pixel Value} - \mu_\text{Noise}}{\sigma}
    \label{eq:snr}
\end{equation}

\newpage
\section{Results}
\begin{figure}[H]
    \centering
    \includegraphics[scale=0.3]{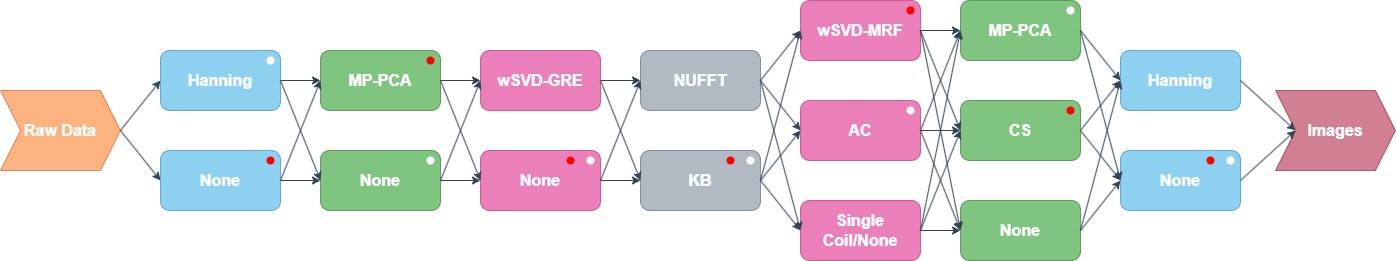}
    \caption{Flowchart of the options in the toolbox. Cyan indicating: filtering, green: denoising, pink: coil combination, and gray: reconstruction.  Red dots indicate the SNR-optimal pipeline path for the MRF dataset, and white dots indicate the SNR-optimal pipeline path for the GRE dataset.}
    \label{flowchart}
\end{figure}

The results are presented as SNR maps of six middle slices from various scans, using different combinations of reconstruction, coil combination, and denoising methods, as seen in the flowchart in Figure \ref{flowchart}. The time used for reconstruction is displayed in \ref{table:timing} for the different reconstruction parameters show an order of magnitude increase from KB to NUFFT.

\begin{table}[h!]
\centering
\begin{tabular}{|c|c|c|c|}
\hline
Method+Length of Spiral/Resolution [s]& 16×16 & 32×32 & 64×64 \\
\hline
KB 33\% spiral & 0.033 & 0.033 & 0.035 \\
\hline
KB 66\% spiral & 0.035 & 0.037 & 0.039 \\
\hline
KB 100\% spiral & 0.038 & 0.039 & 0.040 \\
\hline
NUFFT 33\% spiral & 0.305 & 0.313 & 0.329 \\
\hline
NUFFT 66\% spiral & 0.299 & 0.309 & 0.319 \\
\hline
NUFFT 100\% spiral & 0.299 & 0.308 & 0.332 \\
\hline
\end{tabular}
\caption{Timing used for reconstruction of data with resolutions of 16×16, 32×32, and 64×64 by NUFFT or KB with spiral sampling covering 33\%, 66\%, and 100\% of the k-space}
\label{table:timing}
\end{table}

\subsection{$^{31}$P GRE Data}

\subsubsection{Reconstruction and coil combination for phantom data}

\begin{figure}[H]
    \centering
    \includegraphics[scale=0.42]{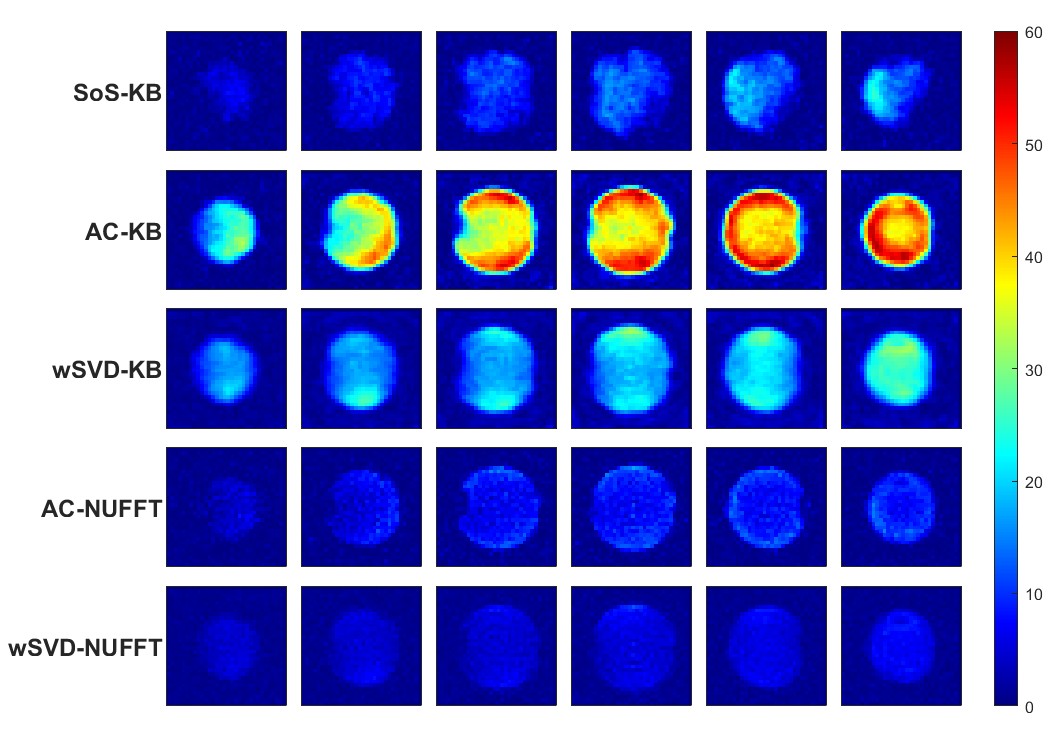}
    \caption{SNR maps of six axial slices of a phosphate phantom, reconstructed using either KB-regridding or NUFFT, with coil combinations performed using SoS, AC, or wSVD.}
    \label{PhantomCoils}
\end{figure}

Figure \ref{PhantomCoils} shows that AC-KB clearly outperforms the other methods. The mean SNR in the region of interest (Table \ref{tab:invivo} ) for AC-KB is more than twice that of wSVD-KB, the second-best method. The NUFFT reconstruction shows a significantly lower SNR, but still preserves the phantom’s geometry clearly. In general, AC outperforms wSVD regardless of the reconstruction method. SoS performs poorly as it shows both low SNR and loss in the phantom geometry with signal dropout at the edges.

\subsubsection{Denoising of phantom data}
\begin{figure}[H]
    \centering
    \includegraphics[scale=0.42]{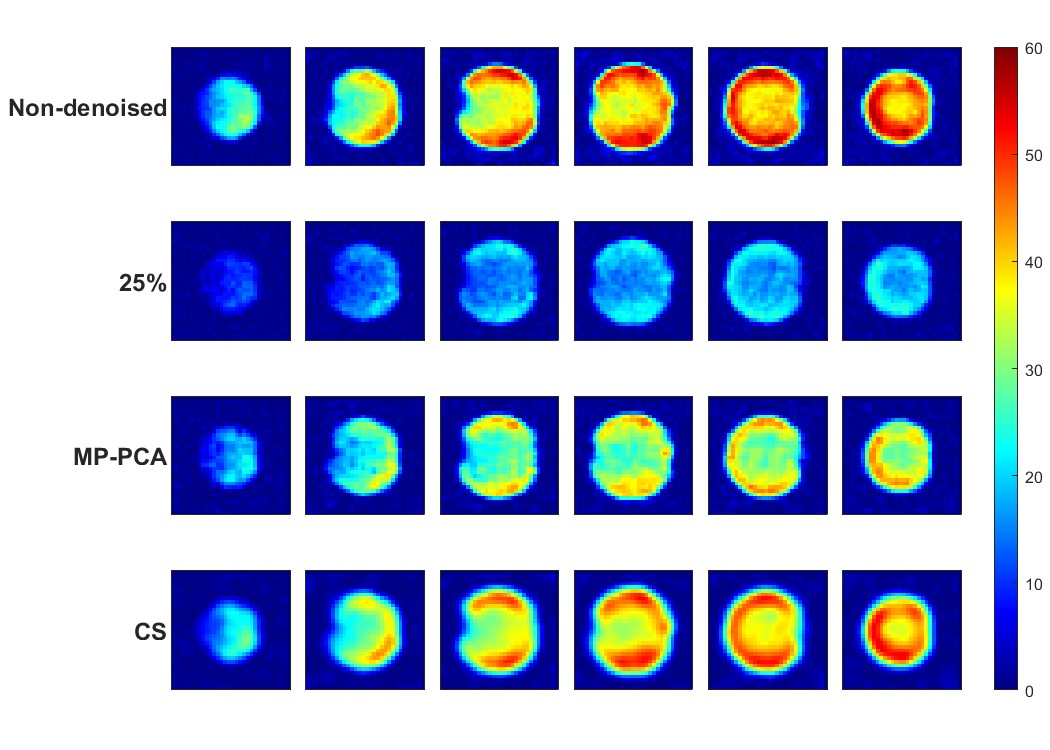}
    \caption{SNR maps of six axial slices of the phantom reconstructed with AC-KB. The maps are presented from top to bottom as follows: the non-denoised and fully-sampled dataset, fourfold downsampled data without denoising, MP-PCA applied on the downsampled data, and CS applied on the downsampled data.}
    \label{PhantomDenoise}
\end{figure}

The denoising was performed on the fourfold undersampled data reconstructed with AC-KB (Figure \ref{PhantomDenoise}). CS gives the highest SNR, which is close to that of the non-denoised fully sampled reference images, with values of 35.2 vs. 38.9, as shown in Table \ref{tab:invivo}. However, CS smooths the SNR maps, while MP-PCA does not. MP-PCA also significantly increases the SNR from the undersampled data.

\subsubsection{Reconstruction and coil combination for in vivo data}
\begin{figure}[H]
    \centering
    \includegraphics[scale=0.42]{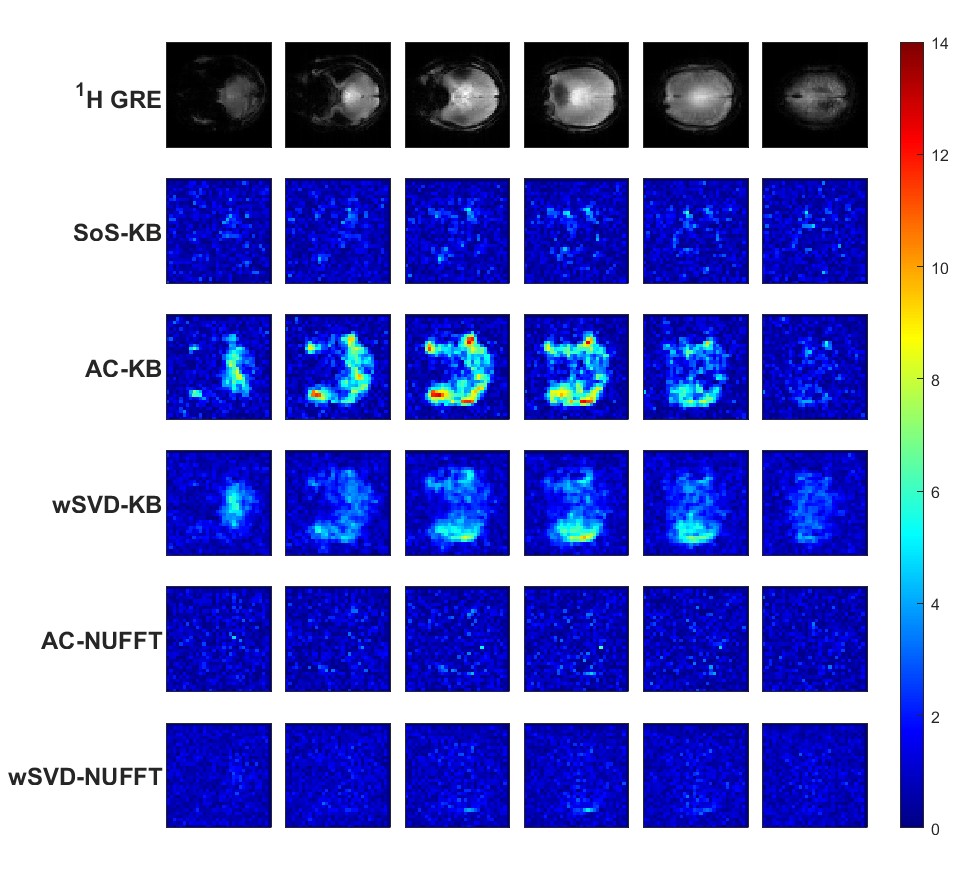}
    \caption{SNR maps of six axial slices from an in vivo GRE scan using 100\% of the k-space data, reconstructed with either KB-regridding or NUFFT, and coil combinations performed using SoS, AC, or wSVD.}
    \label{InVivoCoils}
\end{figure}

In line with the phantom results, AC-KB performs the best, followed by wSVD-KB (Figure \ref{InVivoCoils} ). The other methods fail to preserve the brain geometry. While wSVD-KB provides some signal, it does not retain the structural details as seen with AC-KB. These differences are consistent with the SNR values in Table \ref{tab:invivo}.

\subsubsection{Denoising of in vivo data}
\begin{figure}[H]
    \centering
    \includegraphics[scale=0.77]{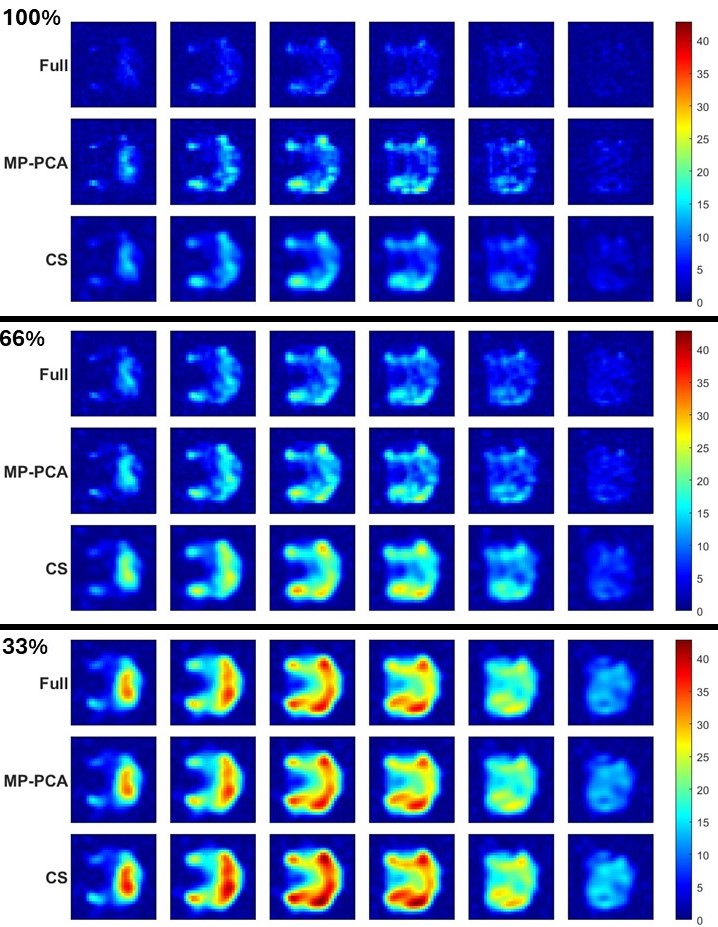}
    \caption{SNR maps of six axial slices from an in vivo GRE phosphorus scan reconstructed with KB-regridding and coils combined with AC. The maps are presented in the following order from top to bottom for each k-space usage level (100\%, 66\%, and 33\%): the non-denoised dataset, MP-PCA applied to the non-denoised dataset, and CS applied to the non-denoised dataset.}
    \label{InvivoDenoise}
\end{figure}

To test denoising of in vivo data, reconstructed images with AC-KB using 100\%, 66\%, and 33\% of the k-space data are presented. An increase in SNR is observed as the proportion of k-space data decreases, demonstrating the effectiveness of filtering. Overall, CS provides a larger SNR boost than MP-PCA (Table \ref{tab:invivo} and Figure \ref{InvivoDenoise}). Similar as seen in the phantom results, MP-PCA preserves the structure while CS smooths the images. This difference diminishes when uses 33 of the k-space data, as the filtering effect from k-space undersampling becomes more dominant.

\subsubsection{Summary of SNR values of 31P GRE images}
\begin{table}[H]
    \centering
    \begin{tabular}{lcc}
        \hline
        Method & Mean & Std \\
        \hline
        \textbf{Reconstruction and coil combination (phantom data)}& & \\
        SoS-KB & 9.6& 4.7\\
        \textbf{AC-KB} & \textbf{38}& \textbf{11}\\
        wSVD-KB & 17& 5.7\\
        AC-NUFFT & 7.5& 3.1\\
        wSVD-NUFFT & 4.7& 1.8\\
        \hline
        \textbf{Denoising (phantom data)}& & \\
        Non-Denoised& 38& 12\\
        Downsampled 4x & 16& 5.3\\
        MP-PCA & 29& 9.4\\
        \textbf{CS} & \textbf{36}& \textbf{10}\\
        \hline
        \textbf{Reconstruction and coil combination (in vivo data)}& & \\
        SoS-KB & 1.1& 0.90\\
        \textbf{AC-KB} & \textbf{3.4}& \textbf{2.5}\\
        wSVD-KB & 2.4& 1.3\\
        AC-NUFFT & 0.92& 0.74\\
        wSVD-NUFFT & 1.0& 0.57\\
        \hline
        \textbf{Denoising (in vivo data)}& & \\
        Non-Denoised 100\%& 3.4& 2.5\\
        CS 100\% & 7.9& 5.0\\
        MP-PCA 100\% & 7.1& 5.0\\
        Non-Denoised 66\%& 8.1& 4.9\\
        CS 66\% & 15& 8.2\\
        MP-PCA 66\% & 10& 6.3\\
        Non-Denoised 33\%& 22& 10\\
        \textbf{CS 33\%} & \textbf{27}& \textbf{13}\\
        MP-PCA 33\% & 22& 9.1 \\

        \hline
    \end{tabular}
    \caption{Mean and standard deviation of the SNR in the phantom and in vivo GRE SNR maps from high signal regions in three middle slices, as shown in supplementary files Figures \ref{noisePhantom} and \ref{noiseiNvivo} where the top performer in every category is boldfaced}
    \label{tab:invivo}
\end{table}

\subsection{$^{31}$P MRF Data}

\subsubsection{Reconstruction and coil combination of in vivo data}
\begin{figure}[H]
    \centering
    \includegraphics[scale=0.42]{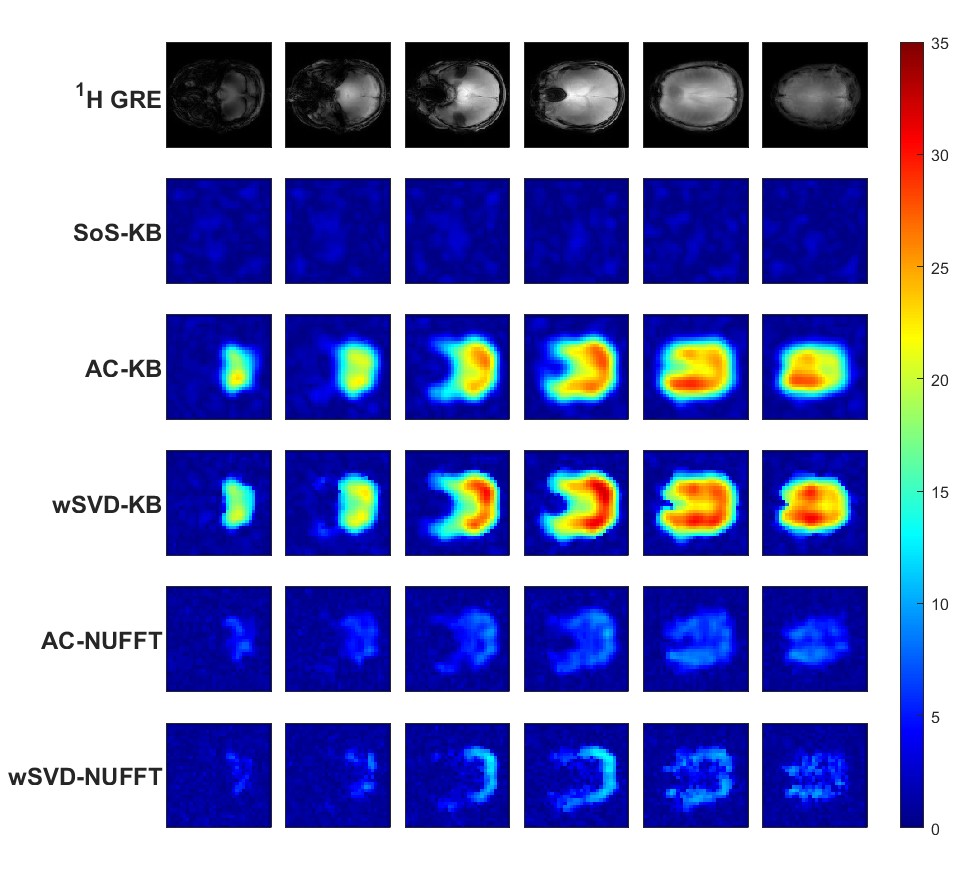}
    \caption{SNR maps of six axial slices of the ATP signal, summed over 400 flip angles (FAs), reconstructed using either KB-regridding or NUFFT, with coil combinations performed using SoS, AC, or wSVD.}
    \label{ATPCoils}
\end{figure}

\begin{figure}[H]
    \centering
    \includegraphics[scale=0.42]{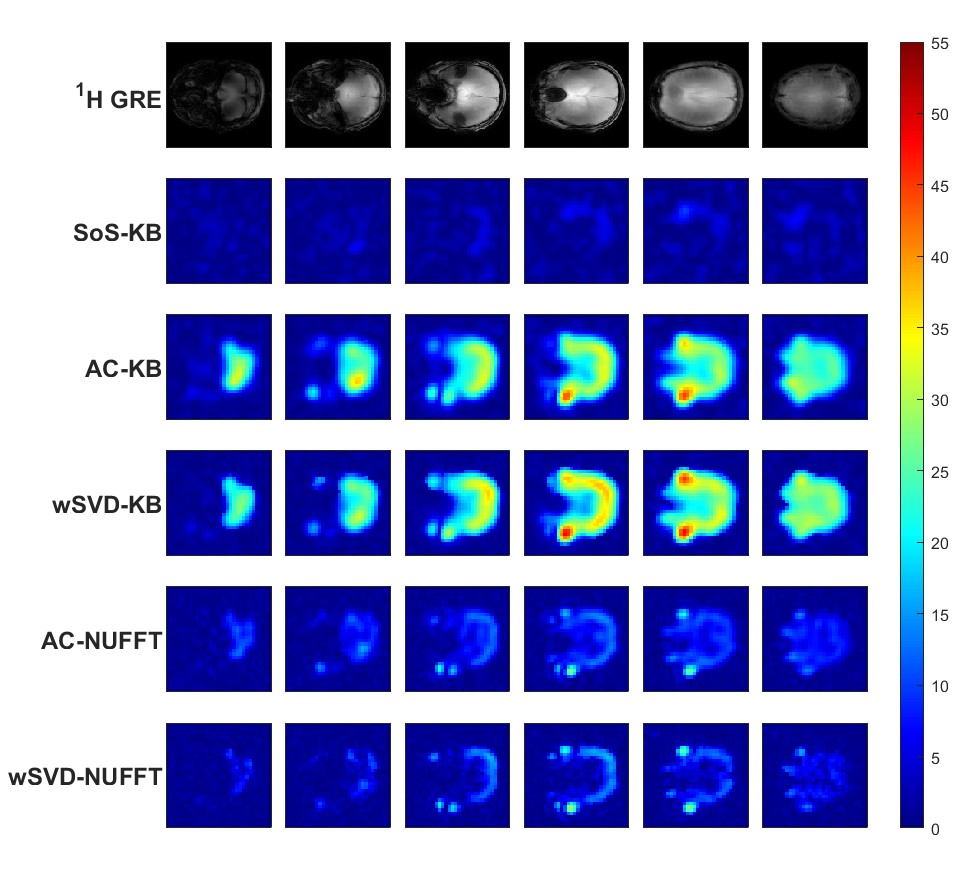}
    \caption{SNR maps of six axial slices of the PCr signal, summed over 200 flip angles (FAs), reconstructed using either KB-regridding or NUFFT, with coil combinations performed using SoS, AC, or wSVD.}
    \label{PCrCoils}
\end{figure}

For the MRF dataset, similar results are observed for both ATP (Figure \ref{ATPCoils} ) and PCr (Figure \ref{PCrCoils}). KB still provides higher SNR than NUFFT, while NUFFT reconstructed images are sharper. In terms of coil combination, SoS produces the lowest SNR maps. Contrary to the GRE data results, wSVD outperforms AC in providing the best SNR maps for MRF data . These visual results are corroborated by mean SNR values in the table (Table \ref{tab:atp_pcr}).

\subsubsection{Denoising of in vivo data} 
\begin{figure}[H]
    \centering
    \includegraphics[scale=0.42]{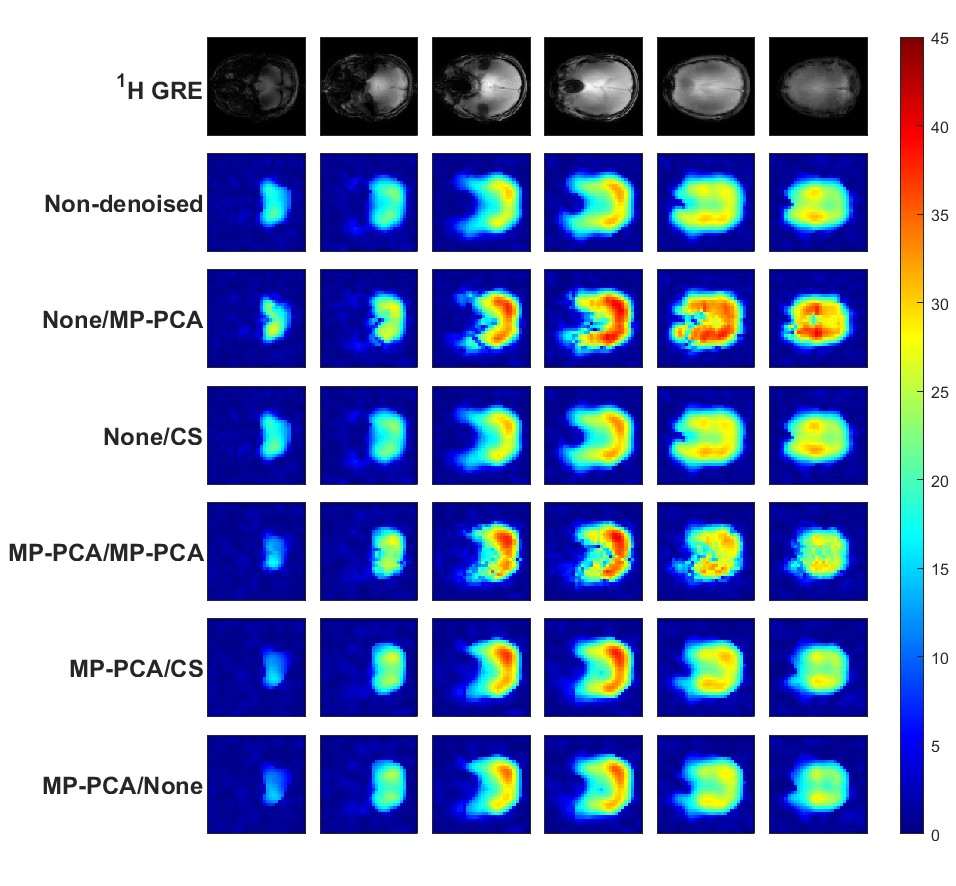}
    \caption{SNR maps of six axial slices from an ATP scan reconstructed with KB and coils combined using wSVD (non-denoised). Denoising methods applied in the time and image domain are stated on the left and right side of the slash, respectively.}
    \label{ATPNoise}
\end{figure}

For the denoising results, ATP and PCr images are examined separately.  Denoising methods applied in the time and image domain are stated on the left and right side of the slash, respectively. In Figure \ref{ATPNoise} and Table \ref{tab:atp_pcr}, denoising methods boost the SNR. However, applying denoising to the time domain alters the shape of the final ATP SNR maps. Applying MP-PCA to the image domain data introduces artifacts, which are particularly visible in the middle slices, albeit offering the highest SNR. CS results in a slight SNR increase. Applying MP-PCA to the time domain data also boosts SNR but alters the shape of SNR maps for ATP but not PCr.

\begin{figure}[H]
    \centering
    \includegraphics[scale=0.42]{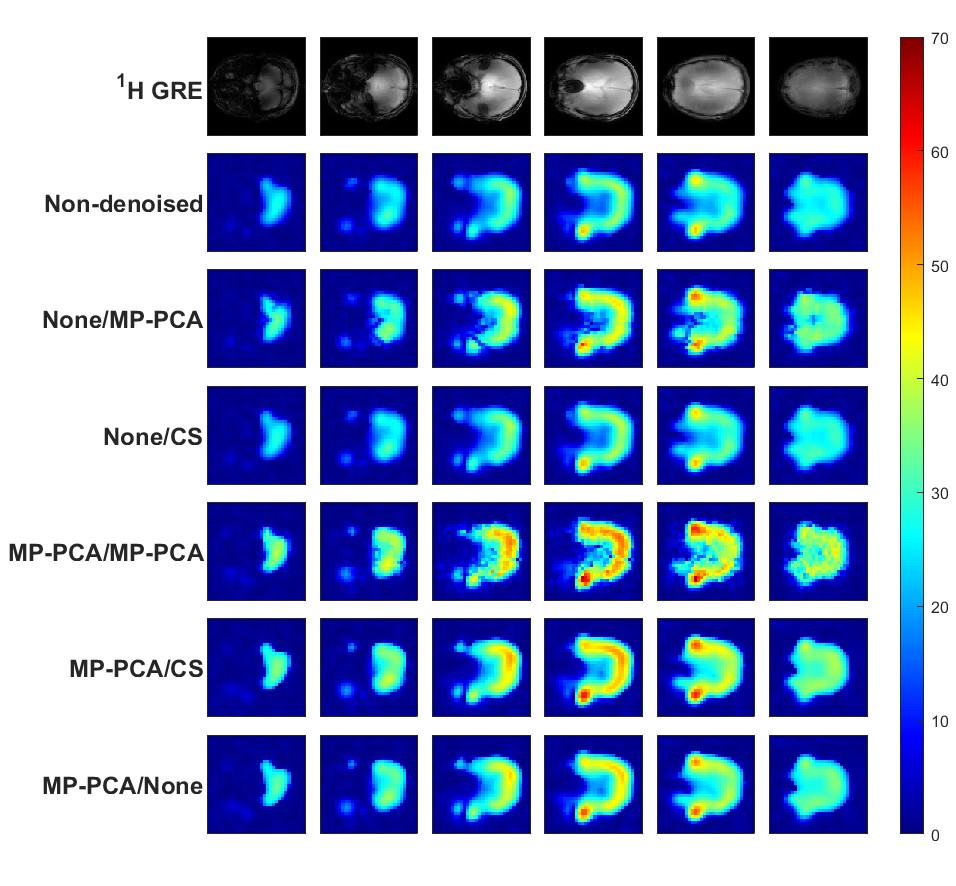}
    \caption{SNR maps of six axial slices from a PCr scan reconstructed with KB and coils combined using wSVD (non-denoised). Denoising methods applied in the time and image domain are stated on the left and right side of the slash, respectively.}
    \label{PCrNoise}
\end{figure}

Denoising results for PCr shown in Figure \ref{PCrNoise}. Applying MP-PCA improves the SNR, however, applying it in the image domain introduces artifacts. CS provides mild increase in SNR without introducing any visible defects in the SNR maps. Among all the combinations, applying MP-PCA for raw data and CS in the image domain provides the highest SNR without artifacts.

\subsubsection{Summary of SNR values of $^{31}P$ MRF images}
\begin{table}[H]
    \centering
    \begin{tabular}{lcc}
        \hline
        Method & Mean & Std \\
        \hline
        \textbf{ATP reconstruction and coil combination} & & \\
        SoS-KB & 0.82& 0.63\\
        AC-KB & 16& 8.1\\
        \textbf{wSVD-KB} & \textbf{18}& \textbf{8.8}\\
        AC-NUFFT & 4.6& 2.5\\
        wSVD-NUFFT & 4.2& 3.0\\
        \hline
        \textbf{ATP denoising} & & \\
        Full & 18& 8.8\\
        \textbf{MP-PCA/None} & \textbf{21}& \textbf{12}\\
        CS/None & 18& 9.2\\
        MP-PCA/MP-PCA & 17& 10\\
        MP-PCA/CS & 17& 9.6\\
        MP-PCA/None & 16& 8.9\\
        \hline
        \textbf{PCr reconstruction and coil combination} & & \\
        SoS-KB & 1.9& 1.7\\
        AC-KB & 21& 9.1\\
        \textbf{wSVD-KB} & \textbf{22}& \textbf{10}\\
        AC-NUFFT & 6.9& 4.0\\
        wSVD-NUFFT & 5.4& 4.7\\
        \hline
        \textbf{PCr denoising}& & \\
        Full & 22& 10\\
        MP-PCA/None & 26& 13\\
        CS/None & 24& 11\\
        \textbf{MP-PCA/MP-PCA} & \textbf{29}& \textbf{15}\\
        MP-PCA/CS & 28& 14\\
        MP-PCA/None & 26& 12\\
        \hline
    \end{tabular}
    \caption{Mean and standard deviation of the SNR in in vivo MRF SNR maps from high signal regions in three middle slices, as shown in Figures \ref{noiseATP} and \ref{NoisePCR} . The method with the highest SNR is highlighted in bold}
    \label{tab:atp_pcr}
\end{table}

\newpage

\section{Discussion}

\subsection{Reconstruction method for low SNR data: KB+FFT vs NUFFT}
\begin{figure}[H]
    \centering
    \includegraphics[scale=0.42]{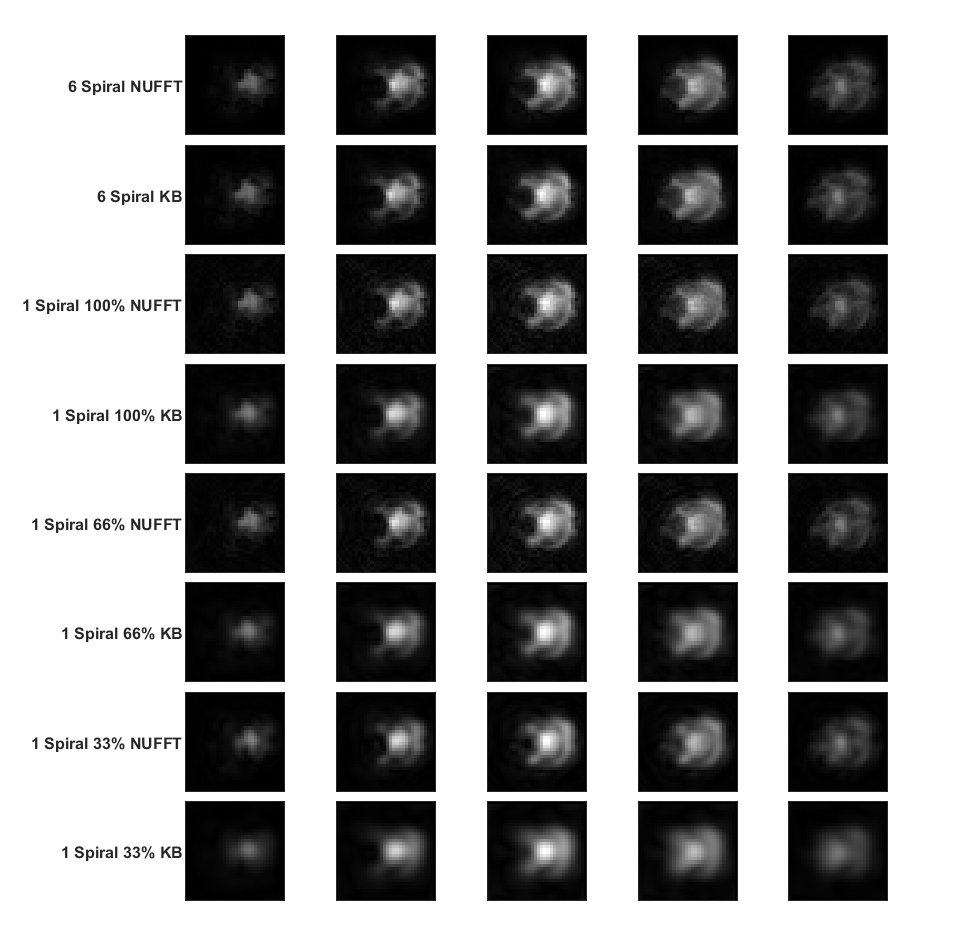}
    \caption{Hydrogen data reconstructed with NUFFT or KB using 6 spirals per slice or 1 while also decreasing the k-space sampling from 100\%, 66\%, and 33\%}
    \label{1HNoNoiseNorm}
\end{figure}

For both 31P datasets investigated in this work, reconstruction using Kaiser-Bessel regridding with FFT offers higher SNR relative to the NUFFT. These results are unexpected and contrary to studies done on $^{1}$H MRI data where NUFFT outperforms KB \cite{nufftBetter,KBbetterNot}. This was investigated in the supplementary files section \textbf{8.4} . It is shown that NUFFT outperforms KB for $^{1}$H GRE data where the SNR of the images are high relative to $^{31}$P GRE images, KB mostly smooths $^{1}$H GRE images(Figure \ref{1HNoNoiseNorm}). To mimic a low SNR data, Gaussian noise was added to the $^{1}$H GRE images to reduce the SNR. This was done by taking the 0.3 times the mean of the first 10 spiral points in the sixth slice, this value was then multiplied with a complex standard normal distribution of the same size as the spiral data and added to it. (Figure \ref{1HNoiseNoCS}). In these high noise level data, NUFFT can no longer reliably reconstruct the data especially in data acquired with one spiral where SNR is even worse, while KB still could reconstruct the images. 
\\
\\
Furthermore, in contrast to the GRE experiments, the number of flip angles (FAs) makes reconstruction time a relevant factor for MRF. NUFFT has a reconstruction time ten times longer than KB—0.4 seconds compared to 0.04 seconds, respectively—which, for 800 FAs, makes a significant difference. 
\\
\\
These results suggest that for low SNR datasets which are common in X-nuclei imaging, KB with FFT is preferable for image reconstruction. Note that NUFFT preserves details in the images, therefore, it can be the choice of method depending on the user's requirements.

\subsection{Data dependent coil combination method}
\begin{figure}[H]
    \centering
    \includegraphics[scale=0.42]{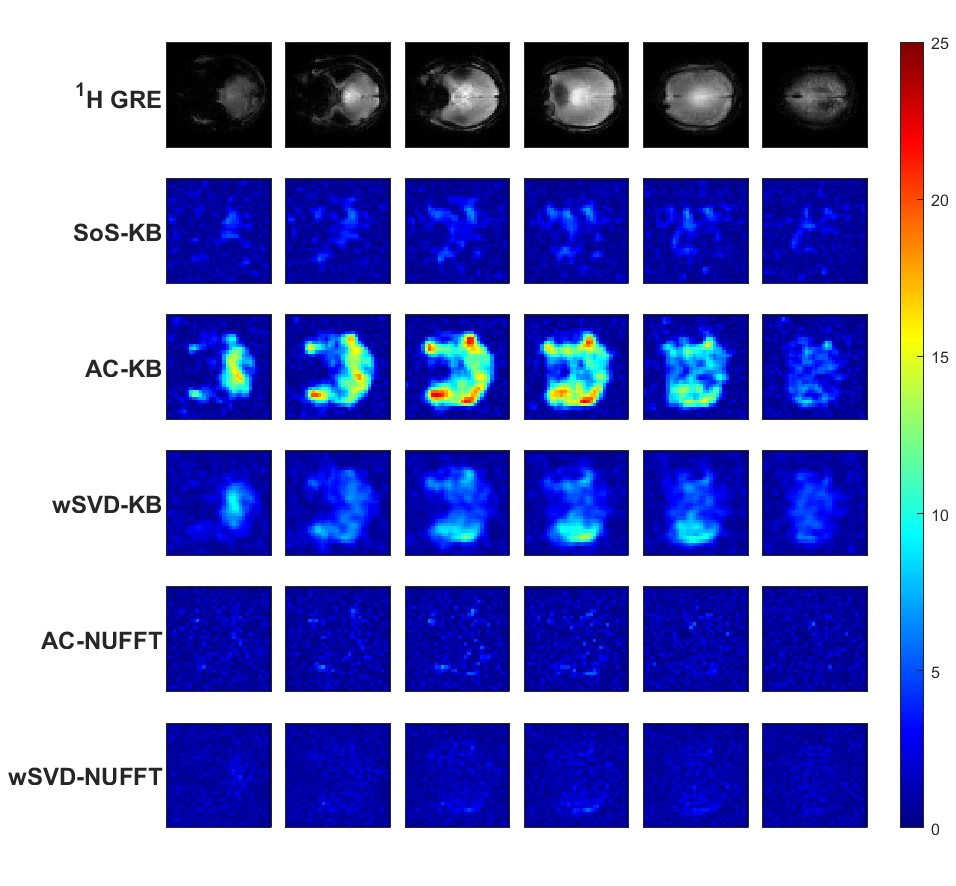}
    \caption{SNR maps of 6 axial slices of a in-vivo 31P GRE scan using 66\% of the k-space reconstructed with KB-regridding or NUFFT and coils combined with SoS, AC or wSVD}
    \label{CC66}
\end{figure}

\begin{figure}[H]
    \centering
    \includegraphics[scale=0.42]{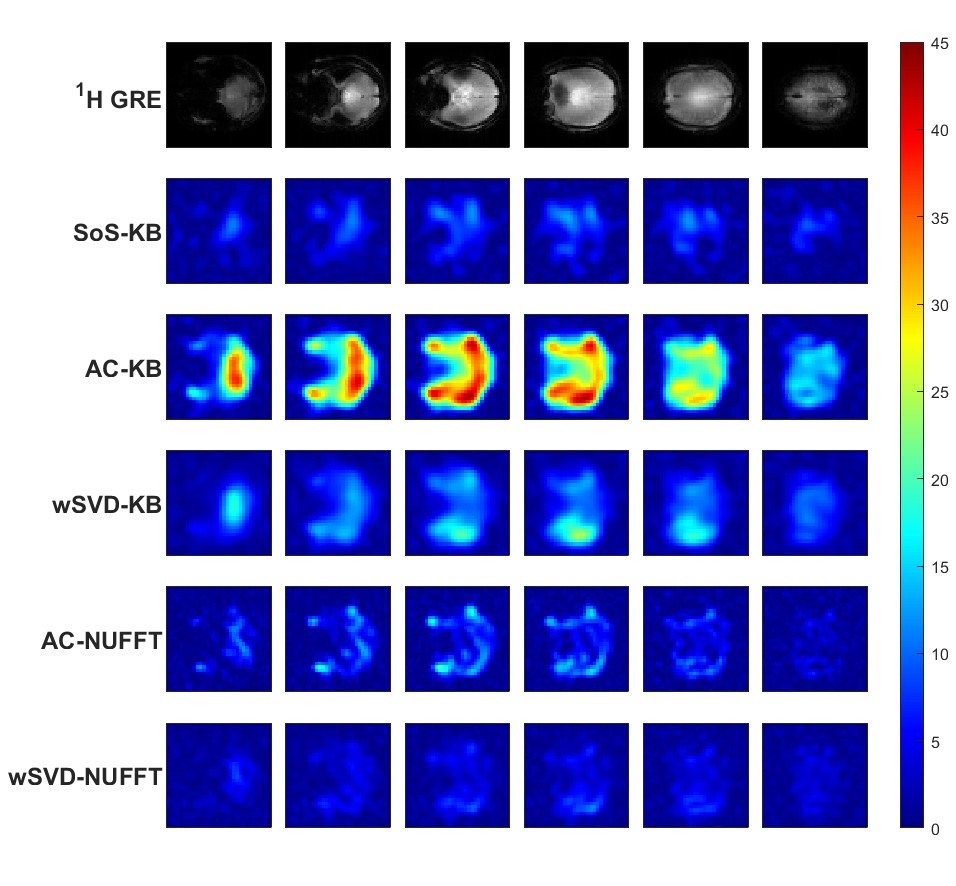}
    \caption{SNR maps of 6 axial slices of a in-vivo 31P GRE scan using 33\% of the k-space reconstructed with KB-regridding or NUFFT and coils combined with SoS, AC or wSVD}
    \label{CC33}
\end{figure}

Both  advanced coil combination methods AC and wSVD demonstrated excellent performance. AC consistently outperforms other methods in GRE scans regardless of the reconstruction technique used.  wSVD outperforms AC for bSSFP MRF data as it does not have the downward gradient of signal that AC does (Fig \ref{PCrCoils} row 3 and 4). These results are consistent for both ATP and PCr images. Indeed, wSVD, a data-driven method, was proposed and recommended in a previous study investigation coil combination methods for 31P MRSI in human torso at 3T. Apodized wSVD methods were evaluated for GRE data, as proposed in algorithm 5 in \ref{wsvdapod} (Figures \ref{CC66} and \ref{CC33}).  A higher SNR with further apodization was observed, but not enough to compete with AC.

\subsection{Performance of denoising methods}
\begin{figure}[H]
    \centering
    \includegraphics[scale=0.42]{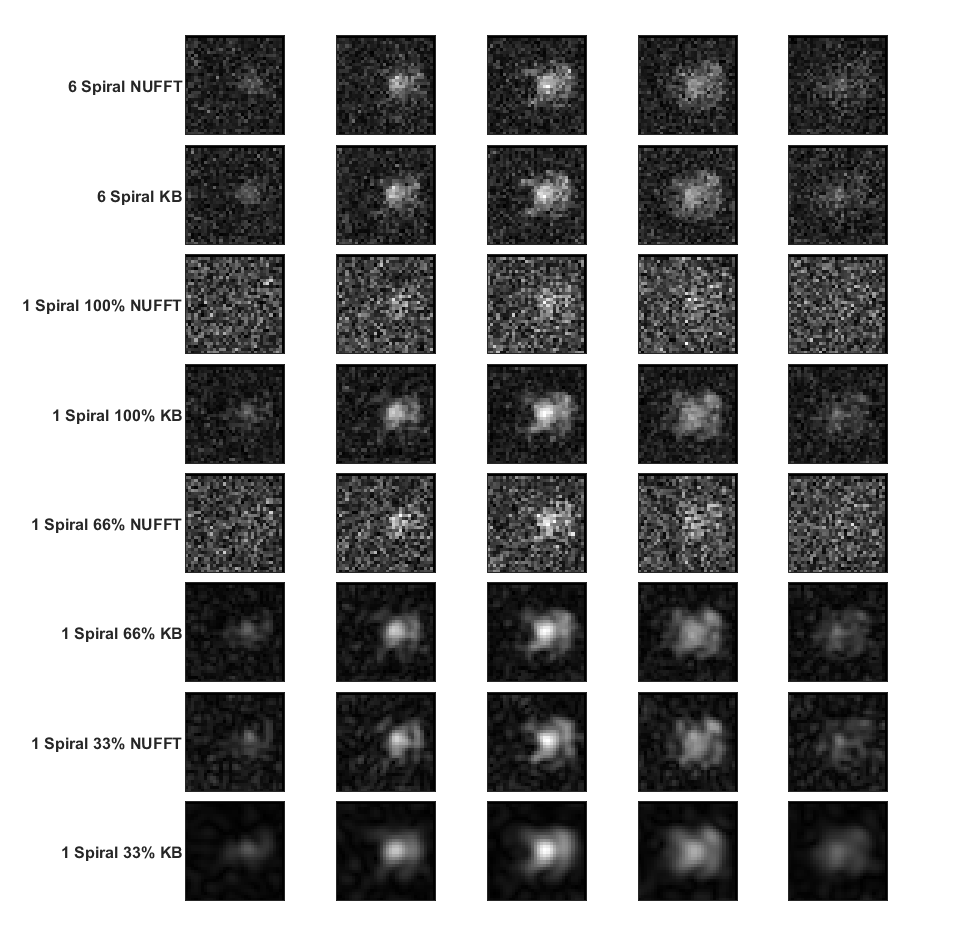}
    \caption{1H GRE MRI images reconstructed with NUFFT or KB using 6 spirals per slice or 1 while also decreasing the k-space sampling from 100\%, 66\%, and 33\%, with added Gaussian noise}
    \label{1HNoiseNoCS}
\end{figure}

CS can successfully remove artifacts in images reconstructed with NUFFT but less effective for denoising in high-noise environments (Figure. \ref{1HNoiseNoCS}). It also has a tendency to smooth images, influenced by the regularization parameter as shown in supplementary files section \textbf{8.3}.
\\
\\
Applying MP-PCA in the image domain generates  artifacts especially for the MRF dataset as the signal for each FA is too low to distinguish signal from noise, while providing an SNR boost simultaneously. When MP-PCA is applied in the time domain for MRF datasets, SNRs of PCr and ATP images were improved, while the structure of the ATP signal degrades.

\newpage
\section{Conclusions}
A toolbox including k-space filtering, coil combination methods (adaptive coil combination and whitened-SVD), , reconstruction methods (Kaiser-Bessel regridding (KB) with FFT and non-uniform FFT), and denoising methods (compressed sensing and MP-PCA) was implemented and investigated for the optimal reconstruction and processing pipeline for fast 31P MR imaging ($^{31}$P GRE and bSSFP-type MRF acqusition). Kaiser-Bessel regridding outperforms NUFFT for the reconstruction of both types of imaging data. Adaptive coil combination outperformed wSVD for GRE data, whereas wSVD showed better performance for MRF data. MP-PCA was effective for GRE data, while CS oversmoothed the SNR maps. For MRF, MP-PCA applied in the image domain introduces artefacts, and CS provided a consistent boost without artefacts. Overall, for 3D $^{31}$P GRE and bSSFP-type MRF data (spatial encoding with a stack of spirals), it is recommended to use Filtering + AC-KB + MP-PCA and MP-PCA + wSVD-KB + CS, respectively.
\\
\\
In conclusion, the choice of method is highly dependent on the type of acquisition and data, and is sensitive to hyperparameters and possible artifacts in the data. Therefore, the optimal pipeline for X-nuclei imaging should be carefully investigated and selected, as the conventional approaches used for hydrogen MRI may not be the most effective. This motivates future investigation to determine an optimal pipeline for other X-nuclei datasets using the toolbox implemented in this work.

\section{References}

\printbibliography[heading=none]

@article{wsvd,
author = {Rodgers, Christopher T. and Robson, Matthew D.},
title = {Receive array magnetic resonance spectroscopy: Whitened singular value decomposition (WSVD) gives optimal Bayesian solution},
journal = {Magnetic Resonance in Medicine},
volume = {63},
number = {4},
pages = {881-891},
doi = {https://doi.org/10.1002/mrm.22230},
year = {2010}
}

@ARTICLE{nufftOG,
  author={Fessler, J.A. and Sutton, B.P.},
  journal={IEEE Transactions on Signal Processing}, 
  title={Nonuniform fast Fourier transforms using min-max interpolation}, 
  year={2003},
  volume={51},
  number={2},
  pages={560-574},
  doi={10.1109/TSP.2002.807005}}

@article{Ma2013,
  author  = {Dan Ma and Vikas Gulani and Nicole Seiberlich and Kecheng Liu and Jeffrey L. Sunshine and Jeffrey L. Duerk and Mark A. Griswold},
  title   = {Magnetic resonance fingerprinting},
  journal = {Nature},
  year    = {2013},
  volume  = {495},
  number  = {7440},
  pages   = {187-192},
  month   = {03},
  doi     = {10.1038/nature11971}
}

@ARTICLE{gridKB,
  author={Beatty, P.J. and Nishimura, D.G. and Pauly, J.M.},
  journal={IEEE Transactions on Medical Imaging}, 
  title={Rapid gridding reconstruction with a minimal oversampling ratio}, 
  year={2005},
  volume={24},
  number={6},
  pages={799-808},
  doi={10.1109/TMI.2005.848376}}

@article{MP1,
  author  = {Jelle Veraart and Els Fieremans and Dmitry S. Novikov},
  title   = {Diffusion MRI noise mapping using random matrix theory},
  journal = {Magnetic Resonance in Medicine},
  year    = {2016},
  volume  = {76},
  number  = {5},
  pages   = {1582-1593},
  doi     = {10.1002/mrm.26059}
}

@article{MP2,
  author  = {Jelle Veraart and others},
  title   = {Denoising of diffusion MRI using random matrix theory},
  journal = {NeuroImage},
  year    = {2016},
  volume  = {142},
  pages   = {394-406},
  doi     = {10.1016/j.neuroimage.2016.08.016}
}

@article{AC,
  author    = {D. O. Walsh and A. F. Gmitro and M. W. Marcellin},
  title     = {Adaptive reconstruction of phased array MR imagery},
  journal   = {Magnetic Resonance in Medicine},
  year      = {2000},
  volume    = {43},
  number    = {5},
  pages     = {682-690},
  doi       = {10.1002/(sici)1522-2594(200005)43:5<682::aid-mrm10>3.0.co;2-g}
}

@article{cs,
author = {Knoll, Florian and Bredies, Kristian and Pock, Thomas and Stollberger, Rudolf},
title = {Second order total generalized variation (TGV) for MRI},
journal = {Magnetic Resonance in Medicine},
volume = {65},
number = {2},
pages = {480-491},
doi = {https://doi.org/10.1002/mrm.22595},
year = {2011}
}

@article{cancer,
  author    = {Klomp, D. W. and van de Bank, B. L. and Raaijmakers, A. and Korteweg, M. A. and Possanzini, C. and Boer, V. O. and van de Berg, C. A. and van de Bosch, M. A. and Luijten, P. R.},
  title     = {31P MRSI and 1H MRS at 7 T: initial results in human breast cancer},
  journal   = {NMR in Biomedicine},
  year      = {2011},
  volume    = {24},
  number    = {10},
  pages     = {1337--1342},
  doi       = {10.1002/nbm.1696},
  url       = {https://doi.org/10.1002/nbm.1696}
}

@article{bipolarity,
  author    = {Deicken, R. F. and Fein, G. and Weiner, M. W.},
  title     = {Abnormal frontal lobe phosphorous metabolism in bipolar disorder},
  journal   = {The American Journal of Psychiatry},
  year      = {1995},
  volume    = {152},
  number    = {6},
  pages     = {915--918},
  doi       = {10.1176/ajp.152.6.915},
  url       = {https://doi.org/10.1176/ajp.152.6.915}
}

@article{xnucl,
title = {The state-of-the-art and emerging design approaches of double-tuned RF coils for X-nuclei, brain MR imaging and spectroscopy: A review},
journal = {Magnetic Resonance Imaging},
volume = {72},
pages = {103-116},
year = {2020},
issn = {0730-725X},
doi = {https://doi.org/10.1016/j.mri.2020.07.003},
url = {https://www.sciencedirect.com/science/article/pii/S0730725X20302368},
author = {Chang-Hoon Choi and Suk-Min Hong and Jörg Felder and N. Jon Shah},
}

@article{AC1lit,
title = {Retrospective correction of bias in diffusion tensor imaging arising from coil combination mode},
journal = {Magnetic Resonance Imaging},
volume = {37},
pages = {203-208},
year = {2017},
issn = {0730-725X},
doi = {https://doi.org/10.1016/j.mri.2016.12.004},
author = {Ken Sakaie and Mark Lowe}
}

@article{usesWSVD1,
  author    = {Hu, W. and Liu, H. and Chen, D. and Qiu, T. and Sun, H. and Xiong, C. and Lin, J. and Guo, D. and Chen, H. and Qu, X.},
  title     = {Coil Combination of Multichannel Single Voxel Magnetic Resonance Spectroscopy with Repeatedly Sampled In Vivo Data},
  journal   = {Molecules},
  year      = {2021},
  month     = {6},
  volume    = {26},
  number    = {13},
  pages     = {3896},
  doi       = {10.3390/molecules26133896}
}

@article{mostlikely,
  author    = {Rodgers, Christopher T. and Robson, Matthew D.},
  title     = {Coil combination for receive array spectroscopy: Are data-driven methods superior to methods using computed field maps?},
  journal   = {Magnetic Resonance in Medicine},
  year      = {2016},
  volume    = {75},
  number    = {2},
  pages     = {473--487},
  doi       = {10.1002/mrm.25618},
  url       = {https://doi.org/10.1002/mrm.25618}
}

@article{mostlikely++,
author = {Mallikourti, Vasiliki and Cheung, Sai and Gagliardi, Tanja and Masannat, Yazan and Heys, Steven and He, Jiabao},
year = {2019},
month = {06},
pages = {},
title = {Optimal Phased-Array Signal Combination For Polyunsaturated Fatty Acids Measurement In Breast Cancer Using Multiple Quantum Coherence MR Spectroscopy At 3T OPEN},
volume = {9},
journal = {Scientific Reports},
doi = {10.1038/s41598-019-45710-1}
}

@article{KBbetterNot,
  author    = {Fessler, Jeffrey A.},
  title     = {On NUFFT-based gridding for non-Cartesian MRI},
  journal   = {Journal of Magnetic Resonance},
  year      = {2007},
  volume    = {188},
  number    = {2},
  pages     = {191--195},
  doi       = {10.1016/j.jmr.2007.06.012},
  url       = {https://doi.org/10.1016/j.jmr.2007.06.012}
}

@article{nufftBetter,
  author    = {Song, J. and Liu, Y. and Gewalt, S. L. and Cofer, G. and Johnson, G. A. and Liu, Q. H.},
  title     = {Least-square NUFFT methods applied to 2-D and 3-D radially encoded MR image reconstruction},
  journal   = {IEEE Transactions on Biomedical Engineering},
  year      = {2009},
  volume    = {56},
  number    = {4},
  pages     = {1134--1142},
  doi       = {10.1109/TBME.2009.2012721},
  url       = {https://doi.org/10.1109/TBME.2009.2012721}
}

@article{DL1,
  author    = {Kaur, Amandeep and Dong, Guanfang},
  title     = {A Complete Review on Image Denoising Techniques for Medical Images},
  journal   = {Neural Processing Letters},
  year      = {2023},
  volume    = {55},
  number    = {6},
  pages     = {7807--7850},
  doi       = {10.1007/s11063-023-11286-1},
  url       = {https://doi.org/10.1007/s11063-023-11286-1},
  issn      = {1573-773X}
}

@inproceedings{DL2,
  author    = {Koppers, Simon and Coussoux, Edouard and Romanzetti, Sandro and Reetz, Kathrin and Merhof, Dorit},
  editor    = {Handels, Heinz and Deserno, Thomas M. and Maier, Andreas and Maier-Hein, Klaus Hermann and Palm, Christoph and Tolxdorff, Thomas},
  title     = {Sodium Image Denoising Based on a Convolutional Denoising Autoencoder},
  booktitle = {Bildverarbeitung f{\"u}r die Medizin 2019},
  year      = {2019},
  publisher = {Springer Fachmedien Wiesbaden},
  address   = {Wiesbaden},
  pages     = {98--103},
  isbn      = {978-3-658-25326-4}
}

@inproceedings{LLR,
  author    = {Moeller, S. and Weingartner, S. and Akcakaya, M.},
  title     = {Multi-scale locally low-rank noise reduction for high-resolution dynamic quantitative cardiac MRI},
  booktitle = {Annual International Conference of the IEEE Engineering in Medicine and Biology Society. IEEE Engineering in Medicine and Biology Society. Annual International Conference},
  year      = {2017},
  pages     = {1473--1476},
  doi       = {10.1109/EMBC.2017.8037113},
  url       = {https://doi.org/10.1109/EMBC.2017.8037113}
}

@article{MP-PCA,
author = {Does, Mark D. and Olesen, Jonas Lynge and Harkins, Kevin D. and Serradas-Duarte, Teresa and Gochberg, Daniel F. and Jespersen, Sune N. and Shemesh, Noam},
title = {Evaluation of principal component analysis image denoising on multi-exponential MRI relaxometry},
journal = {Magnetic Resonance in Medicine},
volume = {81},
number = {6},
pages = {3503-3514},
doi = {https://doi.org/10.1002/mrm.27658},
year = {2019}
}

@article{CSFDA,
  author    = {Ye, Jong Chul},
  title     = {Compressed sensing MRI: a review from signal processing perspective},
  journal   = {BMC Biomedical Engineering},
  year      = {2019},
  volume    = {1},
  number    = {1},
  pages     = {8},
  doi       = {10.1186/s42490-019-0006-z},
  issn      = {2524-4426}
}

@article{Lustig2007,
  author    = {Lustig, Michael and Donoho, David and Pauly, John M.},
  title     = {Sparse MRI: The application of compressed sensing for rapid MR imaging},
  journal   = {Magnetic Resonance in Medicine},
  year      = {2007},
  volume    = {58},
  number    = {6},
  pages     = {1182--1195},
  doi       = {10.1002/mrm.21391},
  url       = {https://doi.org/10.1002/mrm.21391}
}

@article{36,
    author = {Rasche, V. and Proksa, R. and Sinkus, R. and Börnert, P. and Eggers, H.},
    title = {Resampling of data between arbitrary grids using convolution interpolation},
    journal = {IEEE Transactions on Medical Imaging},
    volume = {18},
    number = {5},
    pages = {385--392},
    year = {1999},
    doi = {10.1109/42.774166},
    url = {https://doi.org/10.1109/42.774166}
}

@article{37,
author = {Aurenhammer, Franz},
title = {Voronoi diagrams—a survey of a fundamental geometric data structure},
year = {1991},
issue_date = {Sept. 1991},
publisher = {Association for Computing Machinery},
address = {New York, NY, USA},
volume = {23},
number = {3},
issn = {0360-0300},
doi = {10.1145/116873.116880},
journal = {ACM Comput. Surv.},
month = {sep},
pages = {345–405},
numpages = {61}
}

@misc{38,
    author = {Meng Sang Ong},
    title = {Arbitrary Square Bounded Voronoi Diagram},
    year = {2024},
    howpublished = {\url{https://www.mathworks.com/matlabcentral/fileexchange/30353-arbitrary-square-bounded-voronoi-diagram}},
    note = {MATLAB Central File Exchange. Retrieved January 19, 2024}
}

@article{sodiumNufft1,
  title={Sodium (23Na) ultra-short echo time imaging in the human brain using a 3D-Cones trajectory},
  author={Riemer, F. and Solanky, B.S. and Stehning, C. and others},
  journal={Magnetic Resonance Materials in Physics, Biology and Medicine},
  volume={27},
  number={1},
  pages={35--46},
  year={2014},
  publisher={Springer},
  doi={10.1007/s10334-013-0395-2}
}

@article{sodiumRecon2,
  title={Sodium MRI: methods and applications},
  author={Madelin, Guillaume and Lee, Jin-Suck and Regatte, Ravinder R. and Jerschow, Alexej},
  journal={Progress in Nuclear Magnetic Resonance Spectroscopy},
  volume={79},
  pages={14--47},
  year={2014},
  publisher={Elsevier},
  doi={10.1016/j.pnmrs.2014.02.001}
}

@article{mrf2,
  title={Magnetic resonance fingerprinting: an overview},
  author={Tippareddy, Chandrasekhar and Zhao, Wei and Sunshine, Jeffrey L. and others},
  journal={European Journal of Nuclear Medicine and Molecular Imaging},
  volume={48},
  pages={4189--4200},
  year={2021},
  publisher={Springer},
  doi={10.1007/s00259-021-05384-2}
}

@article{DLMRF,
author = {Hsieh, Jean J. L. and Svalbe, Imants},
title = {Magnetic resonance fingerprinting: from evolution to clinical applications},
journal = {Journal of Medical Radiation Sciences},
volume = {67},
number = {4},
pages = {333-344},
doi = {https://doi.org/10.1002/jmrs.413},
year = {2020}
}

@article{CSsodium1,
author = {Chen, Qingping and Shah, N. Jon and Worthoff, Wieland A.},
title = {Compressed Sensing in Sodium Magnetic Resonance Imaging: Techniques, Applications, and Future Prospects},
journal = {Journal of Magnetic Resonance Imaging},
volume = {55},
number = {5},
pages = {1340-1356},
doi = {https://doi.org/10.1002/jmri.28029},
year = {2022}
}

@ARTICLE{cssodium2,
  author={Speidel, T. and Metze, P. and Rasche, V.},
  journal={IEEE Transactions on Medical Imaging}, 
  title={Efficient 3D Low-Discrepancy  ${k}$ -Space Sampling Using Highly Adaptable Seiffert Spirals}, 
  year={2019},
  volume={38},
  number={8},
  pages={1833-1840},
  doi={10.1109/TMI.2018.2888695}}

@misc{markB1,
      title={Fast 3D 31P B1+ mapping with a weighted stack of spiral trajectory at 7 Tesla}, 
      author={Mark Widmaier and Antonia Kaiser and Salome Baup and Daniel Wenz and Katarzyna Pierzchala and Ying Xiao and Zhiwei Huang and Yun Jiang and Lijing Xin},
      year={2024},
      eprint={2406.18426},
      archivePrefix={arXiv},
      primaryClass={physics.med-ph},
      url={https://arxiv.org/abs/2406.18426}, 
}

@article{KratzerMRF,
author = {Kratzer, Fabian J. and Flassbeck, Sebastian and Schmitter, Sebastian and Wilferth, Tobias and Magill, Arthur W. and Knowles, Benjamin R. and Platt, Tanja and Bachert, Peter and Ladd, Mark E. and Nagel, Armin M.},
title = {3D sodium (23Na) magnetic resonance fingerprinting for time-efficient relaxometric mapping},
journal = {Magnetic Resonance in Medicine},
volume = {86},
number = {5},
pages = {2412-2425},
doi = {https://doi.org/10.1002/mrm.28873},
year = {2021}
}

@inproceedings{markismrm,
  title={3D Whole Brain Mapping of Creatine Kinase Metabolic Rate Using 31P-MR Fingerprinting},
  author={Widmaier, Mark Stephan and Kaiser, Antonia and Xiao, Ying and Huang, Zhiwei and Jiang, Yun and Lim, Song-I and Wenz, Daniel and Xin, Lijing},
  booktitle={Proceedings of the International Society for Magnetic Resonance in Medicine (ISMRM)},
  year={2024},
  location={Singapour},
  note={Program \#0930}
}

@article{wsvdapod,
  author = {Rodgers, Christopher T. and Robson, Matthew D.},
  title = {Coil combination for receive array spectroscopy: Are data-driven methods superior to methods using computed field maps?},
  journal = {Magnetic Resonance in Medicine},
  volume = {75},
  number = {2},
  pages = {473-487},
  year = {2016},
  doi = {10.1002/mrm.25618},
  url = {https://doi.org/10.1002/mrm.25618}
}

\section{Supplementary files}
\subsection{Signal intensity and noise for SNR calculation}

\begin{figure}[H]
    \centering
    \includegraphics[scale=0.6]{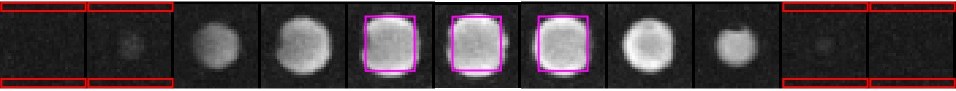}
    \caption{Signal intensity maps of the in phantom scan. The red rectangles indicate the locations used to calculate the standard deviation of the noise for the SNR maps. The pink squares represent the regions from which the SNR mean and standard deviation were calculated in the region of interest.}
    \label{noisePhantom}
\end{figure}

\begin{figure}[H]
    \centering
    \includegraphics[scale=0.6]{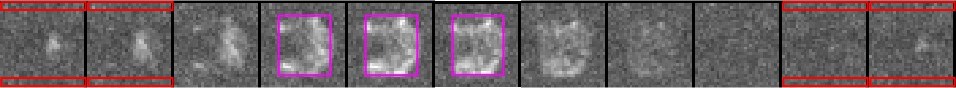}
    \caption{Signal intensity map of the in vivo GRE scan. The red rectangles indicate the locations used to calculate the noise standard deviation in the SNR maps. The pink squares represent the regions from which the SNR maximum and standard deviation were calculated in the region of interest.}
    \label{noiseiNvivo}
\end{figure}

\begin{figure}[H]
    \centering
    \includegraphics[scale=0.6]{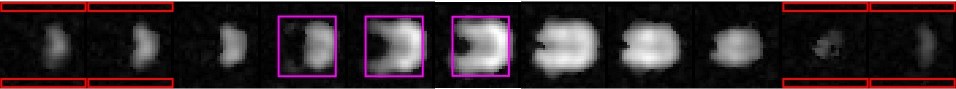}
    \caption{Signal intensity map of the in vivo ATP scan. The red rectangles indicate the locations used to calculate the noise standard deviation in the SNR maps. The pink squares represent the regions from which the SNR maximum and standard deviation were calculated in the region of interest.}
    \label{noiseATP}
\end{figure}

\begin{figure}[H]
    \centering
    \includegraphics[scale=0.6]{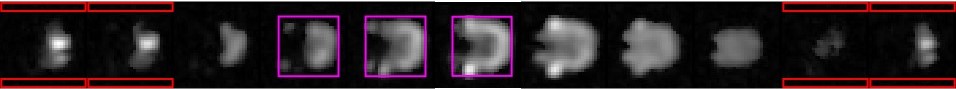}
    \caption{Signal intensity map of the in vivo PCr scan. The red rectangles indicate the locations used to calculate the noise standard deviation in the SNR maps. The pink squares represent the regions from which the SNR maximum and standard deviation were calculated in the region of interest.}
    \label{NoisePCR}
\end{figure}

\subsection{In vivo GRE images: coil combination for different percentages of k-space}

\subsection{Compress sensing experiments}

\begin{figure}[H]
    \centering
    \includegraphics[scale=0.42]{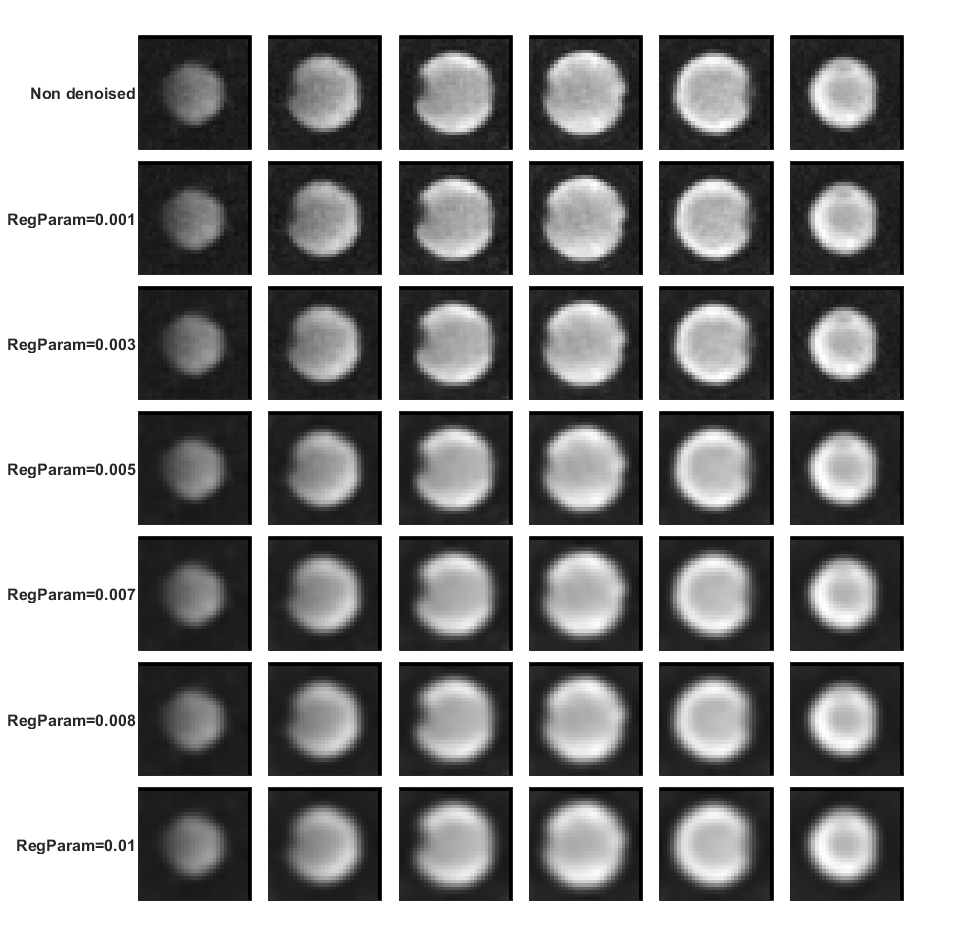}
    \caption{Six middle slices of the phantom reconstructed with KB-AC and CS applied with a regularization parameter from 0.001 to 0.01}
    \label{phlowCSreg}
\end{figure}

\begin{figure}[H]
    \centering
    \includegraphics[scale=0.42]{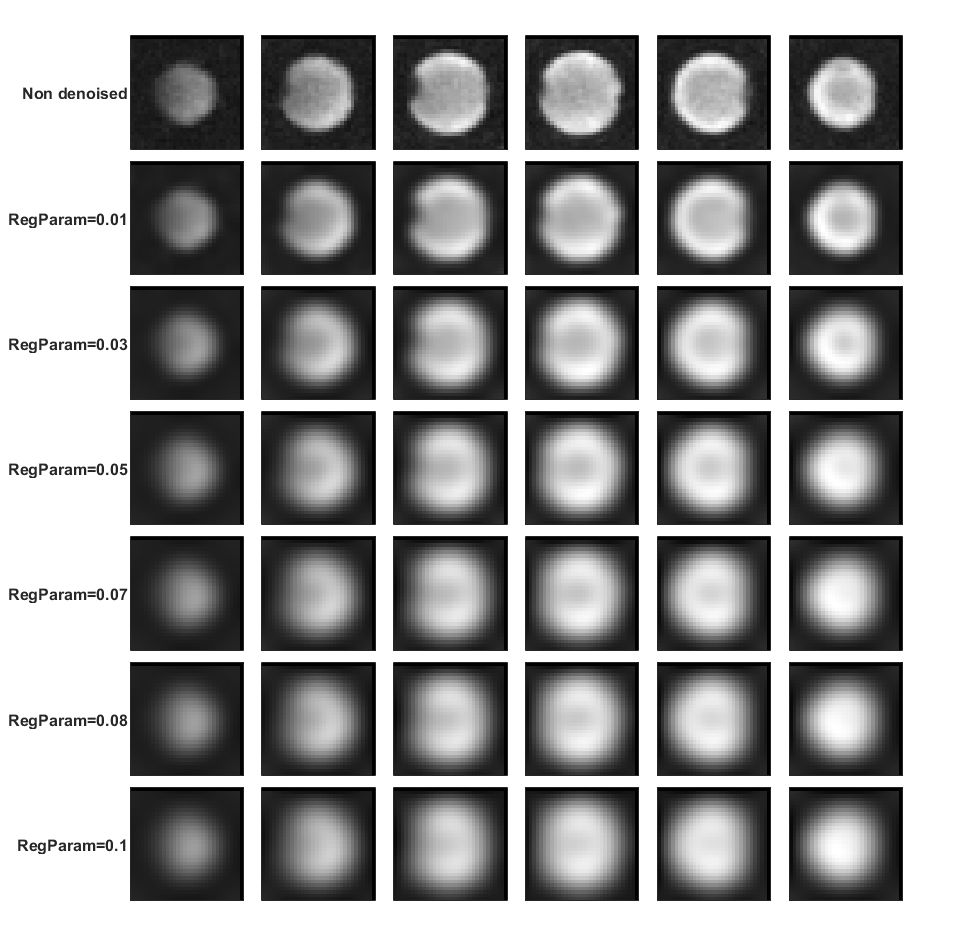}
    \caption{Six middle slices of the phantom reconstructed with KB-AC and CS applied with a regularization parameter from 0.01 to 0.1}
    \label{phhighCSreg}
\end{figure}

\begin{figure}[H]
    \centering
    \includegraphics[scale=0.42]{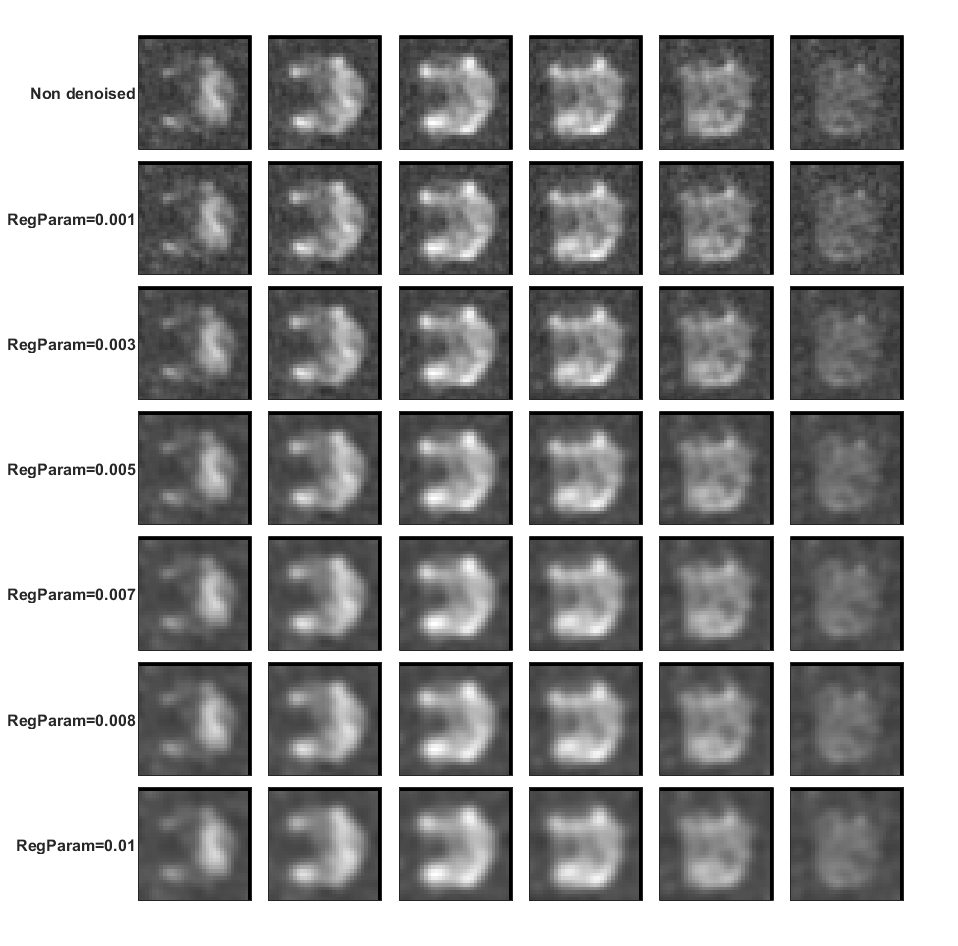}
    \caption{Six middle slices of a in vivo 31P GRE scan reconstructed with KB-AC using 66\% of the k-space and CS applied with a regularization parameter from 0.001 to 0.01}
    \label{66lowCS}
\end{figure}

\begin{figure}[H]
    \centering
    \includegraphics[scale=0.42]{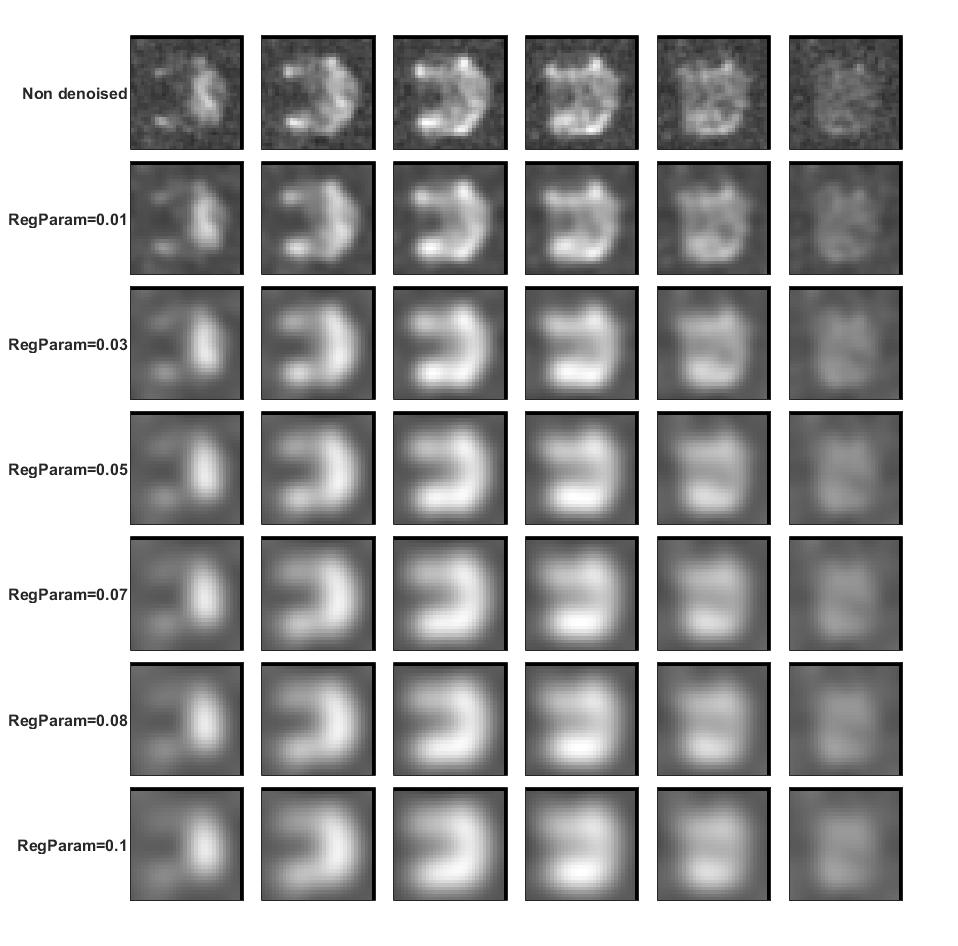}
    \caption{Six middle slices of a in vivo phosphor scan reconstructed with KB-AC using 66\% of the k-space and CS applied with a regularization parameter from 0.01 to 0.1}
    \label{66highCS}
\end{figure}

\begin{figure}[H]
    \centering
    \includegraphics[scale=0.42]{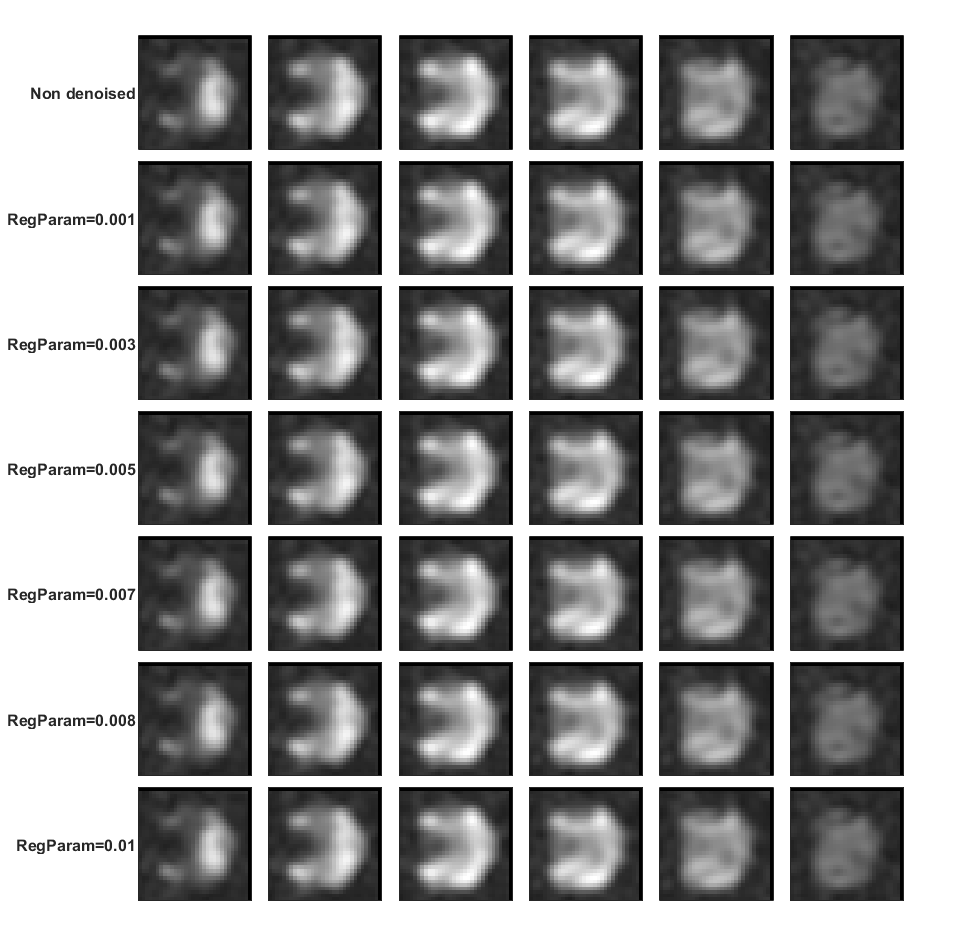}
    \caption{Six middle slices of a in vivo phosphor scan reconstructed with KB-AC using 33\% of the k-space and CS applied with a regularization parameter from 0.001 to 0.01}
    \label{33lowCS}
\end{figure}

\begin{figure}[H]
    \centering
    \includegraphics[scale=0.42]{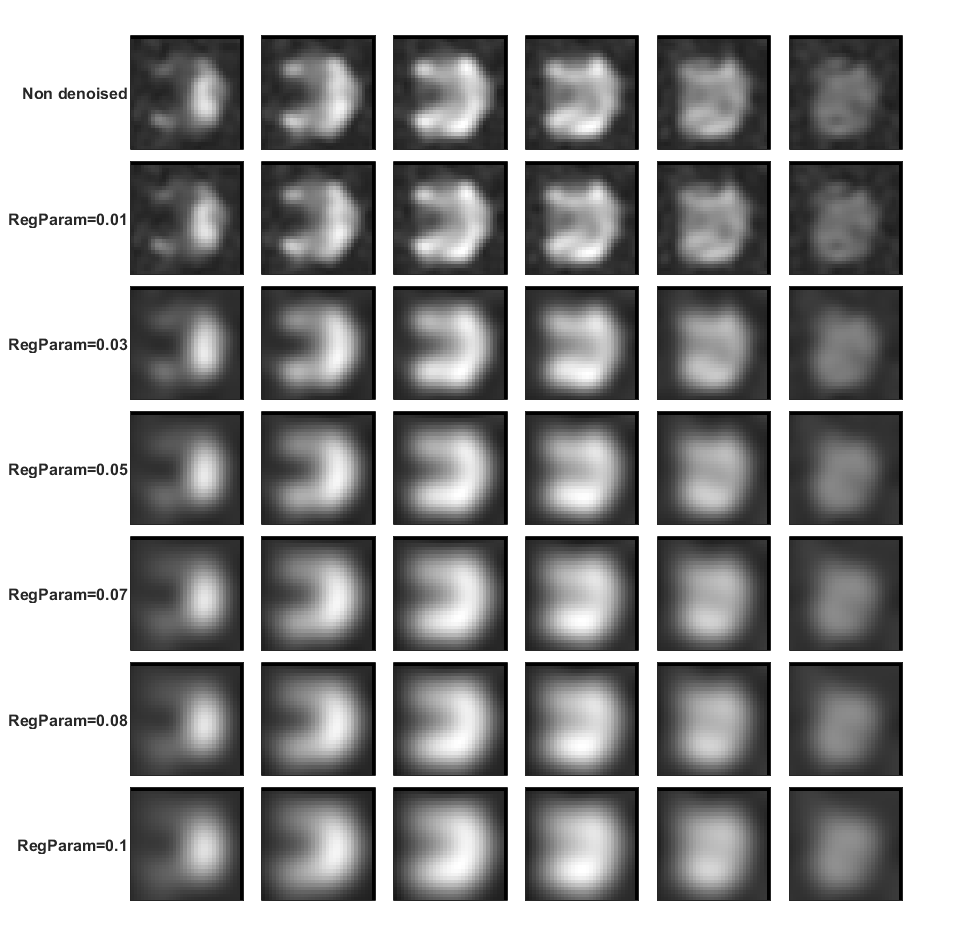}
    \caption{Six middle slices of a in vivo phosphor scan reconstructed with KB-AC using 33\% of the k-space and CS applied with a regularization parameter from 0.01 to 0.1}
    \label{33highCS}
\end{figure}

\subsection{1H MRI results}

\end{document}